\documentclass{sig-alternate-10pt}
\newcommand{\subparagraph}{}
\newcommand*{\titlefont}{\fontfamily{ppl}\selectfont}  
\newcommand{\AUTHORS}{Authors}
\newcommand{\NAME}{{\titlefont{Cell\-Scope}}} 
\newcommand{\TITLE}{Fast and Accurate Performance Analysis \\of LTE Radio Access Networks}
\newcommand{\KEYWORDS}{}
\newcommand{\CONFERENCE}{}
\newcommand{\PAGENUMBERS}{yes}       
\newcommand{\COLOR}{yes}
\newcommand{\COMMENTS}{no}

\usepackage{ifthen, multirow}
\usepackage{fancyvrb}
\usepackage[T1]{fontenc}
\usepackage[utf8]{inputenc}
\usepackage{mathptmx}
\usepackage{bm}
\usepackage[scaled=0.92]{helvet}
\usepackage{inconsolata}
\usepackage[all]{nowidow}

\usepackage[small,compact]{titlesec}              

\usepackage[font=small, labelfont=bf]{caption}
\usepackage{subcaption}

\usepackage{fancyhdr}

\ifthenelse{\equal{\PAGENUMBERS}{yes}}{%
  \pagestyle{plain}
}{%
  \pagestyle{empty}
}

\usepackage[numbers,sort&compress]{natbib}

\usepackage{enumitem}
\setlist{itemsep=0pt,parsep=0pt,topsep=0pt}             

\usepackage{booktabs}
\usepackage{color}
\usepackage{colortbl}
\usepackage{float}                           

\usepackage{graphicx}

\definecolor{placeholderbg}{rgb}{0.85,0.85,0.85}

\usepackage{url}
\ifthenelse{\equal{\COLOR}{yes}}{%
    \usepackage[colorlinks,citecolor=magenta,urlcolor=blue,linkcolor=blue]{hyperref}
}{%
  \usepackage[pdfborder={0 0 0}]{hyperref}
}
\hypersetup{%
pdfauthor = {\AUTHORS},
pdftitle = {\TITLE},
pdfsubject = {\CONFERENCE},
pdfkeywords = {\KEYWORDS},
bookmarksopen = {true}
}
\usepackage{cleveref}
\crefformat{section}{\S#2#1#3}

\usepackage{mathtools}
\usepackage{amssymb}
\usepackage{pifont}

\usepackage{listings}
\definecolor{dkgreen}{rgb}{0,0.6,0}
\lstdefinelanguage{scala}{
  morekeywords={abstract,case,catch,class,def,%
    do,else,extends,false,final,finally,%
    for,if,implicit,import,match,mixin,%
    new,null,object,override,package,%
    private,protected,requires,return,sealed,%
    super,this,throw,trait,true,try,%
    type,val,var,while,with,yield,DStream},
  otherkeywords={=>,<-,<\%,<:,>:,\#,@},
  sensitive=true,
  morecomment=[l]{//},
  morecomment=[n]{/*}{*/},
  morestring=[b]",
  morestring=[b]',
  morestring=[b]""",
  commentstyle=\color{dkgreen},
  basicstyle=\footnotesize\ttfamily
}

\usepackage{algorithm}
\usepackage{algpseudocode}

\usepackage{soul}
\soulregister\cite7
\soulregister\ref7
\soulregister\pageref7

\ifthenelse{\equal{\COMMENTS}{yes}}{%
    \setlength{\marginparwidth}{2cm}
    \usepackage[draft]{todonotes}
}{%
    \usepackage[disable]{todonotes}
}%

\usepackage{microtype}  

\begin{document}

\date{}

\title{\TITLE}

\numberofauthors{1}
\author{
    \alignauthor
Anand Padmanabha Iyer\textsuperscript{$\star$}, Ion
Stoica\textsuperscript{$\star$}, Mosharaf Chowdhury\textsuperscript{$\diamond$}, Li
Erran Li\textsuperscript{$\dag$}\\
       \affaddr{\textsuperscript{$\star$}University of California, Berkeley
           \qquad \textsuperscript{$\diamond$}University of Michigan
       \qquad \textsuperscript{$\dag$}Uber Technologies}\\
}

\makeatletter
\def\blfootnote{\xdef\@thefnmark{}\@footnotetext}
\makeatother

\maketitle 

\ifthenelse{\equal{\PAGENUMBERS}{no}}{%
  \thispagestyle{empty}
}

\blfootnote{\textsuperscript{$\dag$}Li Erran Li was involved in this work prior
    to joining Uber.}

\begin{abstract}
An increasing amount of analytics is performed on data that is procured in a
real-time fashion to make real-time decisions. Such tasks include simple
reporting on streams to sophisticated model building. However, the practicality
of such analyses are impeded in several domains because they are faced with a
fundamental trade-off between data collection latency and analysis accuracy. 

In this paper, we study this trade-off in the context of a specific domain,
Cellular Radio Access Networks (RAN). Our choice of this domain is influenced by
its commonalities with several other domains that produce real-time data, our
access to a large live dataset, and their real-time nature and dimensionality
which makes it a natural fit for a popular analysis technique, machine learning
(ML). We find that the latency accuracy trade-off can be resolved using two
broad, general techniques: intelligent data grouping and task formulations that
leverage domain characteristics. Based on this, we present \NAME{}, a system
that addresses this challenge by applying a domain specific formulation and
application of Multi-task Learning (MTL) to RAN performance analysis. It
achieves this goal using three techniques: feature engineering to transform raw
data into effective features, a PCA inspired similarity metric to group data
from geographically nearby base stations sharing performance commonalities, and
a hybrid online-offline model for efficient model updates. Our evaluation of
\NAME{} shows that its accuracy improvements over direct application of ML range
from 2.5$\times$ to 4.4$\times$ while reducing the model update overhead by up
to 4.8$\times$. We have also used \NAME{} to analyze a live LTE consisting of
over 2 million subscribers for a period of over 10 months, where it uncovered
several problems and insights, some of them previously unknown.
\end{abstract}

\section{Introduction} 
\label{sec:introduction}
Big data analytics have seen tremendous growth in the recent past, with
applications in several domains such as social networks, sciences and medicine.
Increasingly, the trend in data analytics have moved towards tasks that operate
on data that is procured in a real-time fashion and produce low latency
decisions. However, such analysis tasks are faced with a fundamental trade-off
between latency and accuracy in several domains. In this paper, we consider this
trade-off in the context of a domain specific use case: performance diagnostics
in cellular Radio Access Networks (RANs). Being the most crucial component in
the cellular network architecture, the performance of RANs (in terms of
throughput, latency, call drops, etc.) is essential for achieving high quality
of experience for end-users.  Although RANs work in tandem with the cellular
core network to provide service, the lion's share of user-facing issues in a
typical cellular network manifest in the RAN~\cite{RANvsCore}. Thus, to achieve
high end-user satisfaction, it is imperative that operators understand the
impacting factors and can diagnose RAN problems quickly.

While RAN technologies have seen tremendous improvements over the past
decade~\cite{3G,4G,LTE}, performance problems are still
prevalent~\cite{PerformanceSigmetrics2013}. Factors impacting RAN performance
include user mobility, skewed traffic pattern, interference, lack of coverage,
unoptimized configuration parameters, inefficient algorithms, equipment
failures, software bugs and protocol
errors~\cite{NetworkDynamicsSigmetrics2014}. Although some of these factors are
present in traditional networks and troubleshooting these networks has received
considerable attention in the
literature~\cite{TroubleshootingSigcomm2007,FailureICAC2004,CorrelationOSDI2004,NetPrintsNSDI2009,PeerpressureOSDI2004},
RAN performance diagnosis brings out a unique challenge: the performance of
multiple base stations exhibit complex temporal and spatial interdependencies
due to the shared radio access media and user mobility.

Existing systems~\cite{xNetHealing2,NPO} for detecting performance problems rely
on monitoring aggregate metrics, such as connection drop rate and throughput per
cell, over minutes-long time windows. Degradation of these metrics trigger
mostly manual---hence, time-consuming and error-prone---root cause analysis.
Furthermore, due to their dependence on aggregate information, these tools
either overlook many performance problems such as temporal spikes leading to
cascading failures or are unable to isolate root causes. The challenges
associated with leveraging just aggregate metrics has led operators to collect
detailed traces from their network~\cite{xNetHealing1} to aid domain experts in
diagnosing RAN problems.

However, the sheer volume of the data and its high dimensionality make the
troubleshooting using human experts and traditional rule-based systems very
hard, if not infeasible~\cite{RulebasedDSC2007}. In this paper, we consider one
natural alternative to these approaches that has been used recently to
troubleshoot other complex systems with considerable success: machine learning
(ML)~\cite{ML}. However, simply applying ML to our problem is not enough. The
desire to troubleshoot RANs as fast as possible exposes the inherent tradeoff
between \textit{latency} and \textit{accuracy} that is shared by many ML
algorithms.

To illustrate this tradeoff, consider the natural solution of building a model
on a per-base station basis. On one hand, if we want to troubleshoot quickly,
the amount of data collected for a given base station may not be enough to learn
an accurate model. On the other hand, if we wait long enough to learn a more
accurate model, this will come at the cost of delaying troubleshooting and the
learned model may not be valid any longer. Another alternative would be to learn
one model over the entire data set.  Unfortunately, since base stations can have
very different characteristics using a single model for all of them can also
result in low accuracy (\cref{sec:motivation}).

In this paper, we present \NAME{}, a system that enables fast and accurate RAN
performance diagnosis by resolving the \textit{latency} and \textit{accuracy}
trade-off using two broad techniques: intelligent data grouping and task
formulations that leverage domain characteristics.  More specifically, \NAME{}
applies Multi-task Learning (MTL)~\cite{MTLNIPS1996,MTLICML1993}, a
state-of-the-art machine learning approach, to RAN troubleshooting. In a
nutshell, MTL learns multiple related models in parallel by leveraging the
commonality between those models. To enable the application of MTL, \NAME{} uses
two techniques. First, it uses \textit{feature engineering} to identify the
relevant features to use for learning.  Second, it uses a PCA based similarity
metric to group base stations that share common features, such as interference
and load.  This is necessary since MTL assumes that the models have some
commonality which is not necessarily the case in our setting, e.g., different
base stations might exhibit different features. Note that while PCA has been
traditionally used to find network anomalies, \NAME{} uses PCA for finding the
common features instead. We note that \textit{the goal of \NAME{} is
    not to apply specific ML algorithms for systems diagnostics, but to propose
    approaches to resolve the latency accuracy trade-off} common in many domains.

To this end, \NAME{} uses MTL to create a hybrid model: an offline
base model that captures common features, and an online per-base
station model that captures the individual features of the base
stations. This hybrid approach allows us to incrementally update the
online model based on the base model. This results in models that are
both accurate and fast to update. Finally, in this approach, finding
anomalies is  equivalent to detecting concept drift~\cite{ConceptDriftSurvey}.

To demonstrate the effectiveness of our proposal, we have built \NAME{} on
Spark~\cite{RDD, MLI13, MLbase13}. Our evaluation shows that \NAME{} is able to achieve accuracy
improvements upto 4.4$\times$ without incurring the latency overhead associated
with normal approaches (\cref{sec:evaluation}). We have also used \NAME{} to
analyze a live LTE network consisting of over 2 million subscribers for a period
of over 10 months. Our analysis reveals several interesting
insights~(\cref{sec:findings}). 

In summary, we make the following contributions:
\begin{itemize} 
    \item We expose the fundamental trade-off between data collection latency
        and analysis accuracy present in many domains, which impedes the
        practicality of applying analytics for decision making on data collected in a real-time
        fashion. We find that this trade-off may be resolved in several domains using two broad approaches:
        intelligent grouping and domain specific formulations.
    \item Based on this insight, we present \NAME{}, a system that applies a domain specific formulation and application of Multi-task
        Learning (MTL) to resolve the \textit{latency} and \textit{accuracy} trade-off in RAN performance
        analysis. It achieve this using three techniques: \textit{feature
        engineering} to transform raw data into effective features, a novel
    \textit{PCA inspired similarity metric} to group data from base stations
    sharing commonalities in performance, and a \textit{hybrid
        online-offline model} for efficient model updates
    (\cref{sec:cellscope}).
    \item We have built \NAME{} on Apache Spark, a big data
        framework. Our evaluation shows that \NAME{}'s accuracy improvements
        range from 2.5$\times$ to 4.4$\times$ while reducing the model update
        overhead by up to 4.8$\times$ (\cref{sec:evaluation}). We have also
        validated \NAME{} by using it to analyze an operational LTE consisting of over
        2 million subscribers for a period of over 10 months. Our analysis uncovered insights which were valuable for operators (\cref{sec:findings}).  
\end{itemize}

\section{Background and Motivation}
\label{sec:motivation}
In this section, we briefly discuss cellular networks, focusing on the LTE
network architecture, protocol procedures and measurement data and then motivate
the problem.

\subsection{LTE Network Primer}
\label{sec:lte_primer}
\begin{figure}[!t]
\centering
\includegraphics[width=\linewidth]{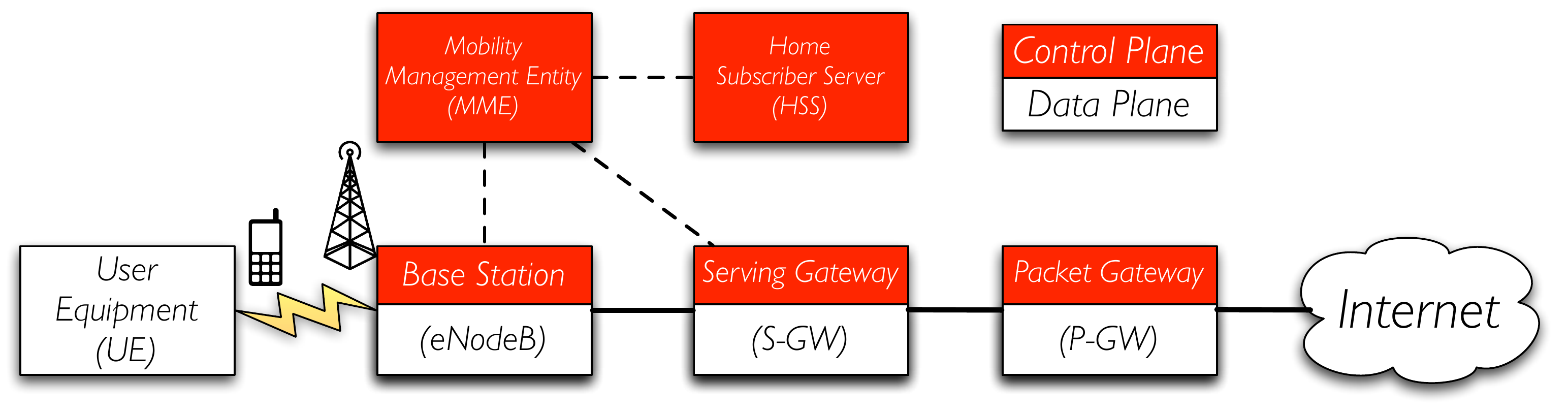}
\caption{LTE network architecture}
\label{fig:LTE-arch}
\end{figure}

LTE networks provide User Equipments (UEs) such as smartphones with Internet
connectivity. When a UE has data to send to or receive from the Internet, it
sets up a communication channel between itself and the Packet Data Network
Gateway (P-GW). This involves message exchanges between the UE and the Mobility
Management Entity (MME). In coordination with the base station (eNodeB), the
Serving Gateway (S-GW), and P-GW, data plane (GTP) tunnels are established
between the base station and the S-GW, and between the S-GW and the P-GW.
Together with the connection between the UE and the base station, the network
establishes a communication channel called EPS bearer (short for bearer). The
entities in the LTE network architecture is shown in figure~\ref{fig:LTE-arch}. 

\begin{figure*}[!t]
\centering
\begin{subfigure}{.33\textwidth}
  \centering
  \includegraphics[width=\linewidth]{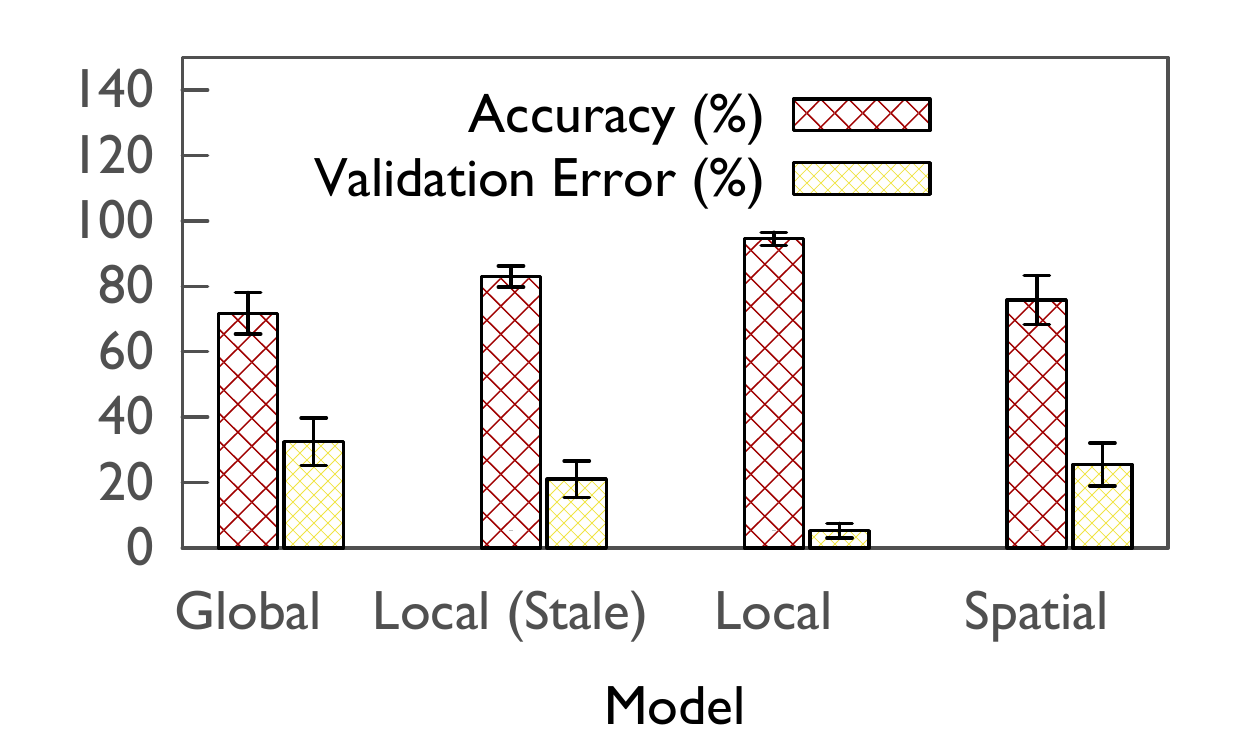}
  \captionof{figure}{Global models are ineffective while spatial partitioning
      ignores performance similarity}
  \label{fig:spatialmodel}
  \end{subfigure}%
\begin{subfigure}{.33\textwidth}
  \centering
  \includegraphics[width=\linewidth]{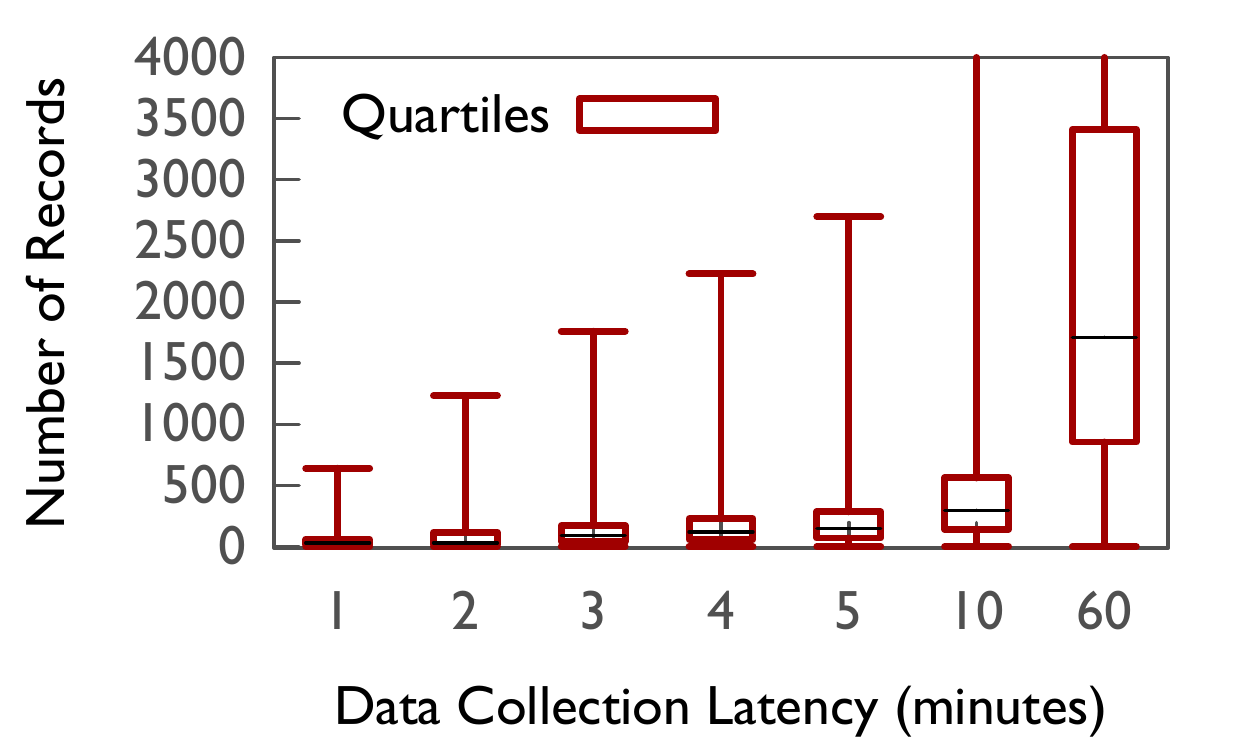}
  \captionof{figure}{Lack of data at low latencies indicates
      the need for grouping/partitoning}
  \label{fig:grouping}
  \end{subfigure}
\begin{subfigure}{.33\textwidth}
  \centering
  \includegraphics[width=\linewidth]{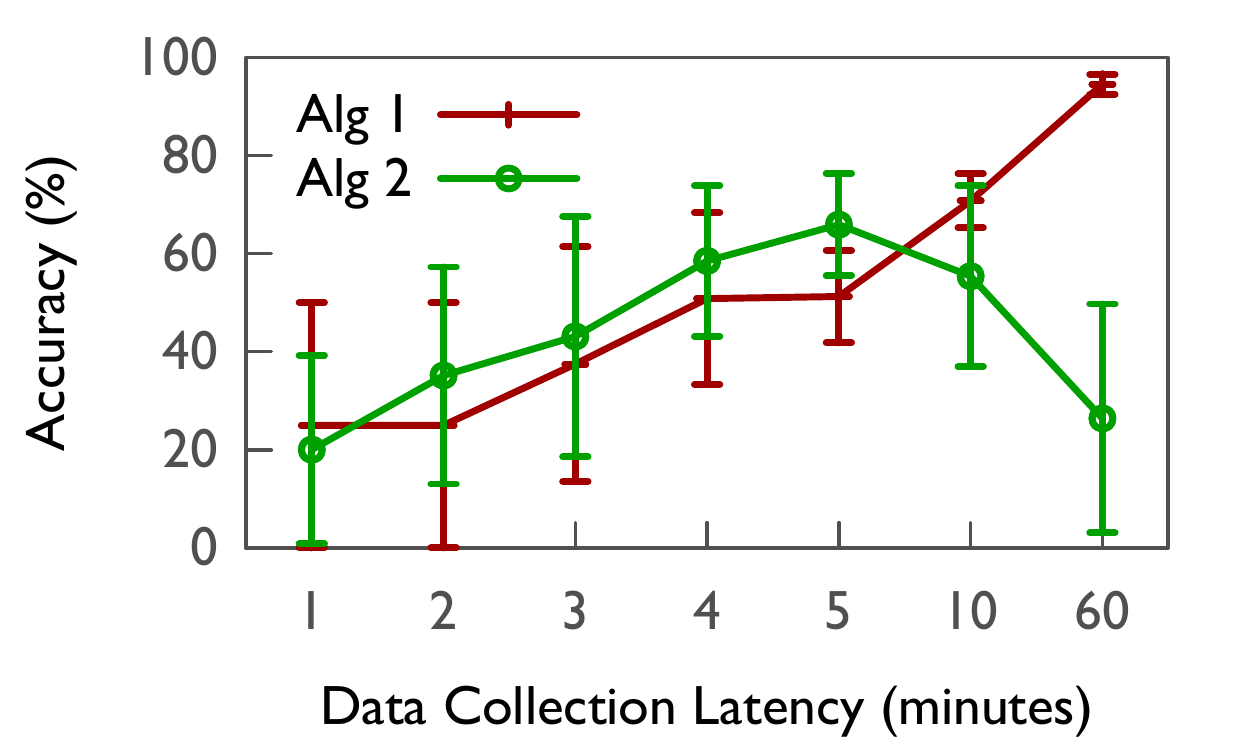}
  \captionof{figure}{Using local models leads to accuracy and/or latency issues}
  \label{fig:latencyissue}
 \end{subfigure}
 \caption{Resolving the latency accuracy trade-off requires domain specific
     optimizations.}
\end{figure*}

For network access and service, entities in the LTE network exchange control
plane messages. A specific sequence of such control plane message exchange is
called a \textit{network procedure}. For example, when a UE powers up, it
initiates an \textit{attach procedure} with the MME which consists of
establishing a radio connection to the base station, authentication and resource
allocation. Thus, each network procedure involves the exchange of several
control plane messages between two or more entities. The specifications for
these are defined by various 3GPP Technical Specification Groups
(TSG)~\cite{3gpp-specs}. 

Network performance degrades and end-user experience is affected when procedure
failures happen. The complex nature of these procedures (due to the multiple
underlying message and entity interactions) make diagnosing problems
challenging. Thus, to aid RAN troubleshooting, operators collect extensive
measurements from their network. These measurements typically consist of
per-procedure information (e.g., attach). To analyze a procedure failure, it is
often useful to look at the associated variables. For instance, a failed
attachment procedure may be diagnosed if the underlying signal strength
information was captured \footnote{Some of the key physical layer parameters
    useful for diagnosis is described in table~\ref{tbl:phyparams}.}.  Hence,
relevant metadata is also captured with procedure information. Since there are
hundreds of procedures in the network and each procedure can have many possible
metadata fields, the collected measurement data contains several hundreds of
fields\footnote{Our dataset consists of almost 400 fields in the measurement
    data, with each field possibly having additional nested information.}.  

\subsection{RAN Troubleshooting Today}
Current RAN network monitoring depends on cell-level aggregate Key Performance
Indicators (KPI). Existing practice is to use performance counters to derive
these KPIs. The derived KPIs are then monitored by domain experts, aggregated
over certain pre-defined time window. Based on domain knowledge and operational
experience, these KPIs are used to determine if service level agreements (SLA)
are met. For instance, an operator may have designed the network to have no more
than 0.5\% call drops in a 10 minute window. When a KPI that is being monitored
crosses the threshold, an alarm is raised and a ticket created. This ticket is
then handled by experts who investigate the cause of the problem, often
manually. Several commercial solutions exists~\cite{WNG, NPO,
    MotiveSC,xNetHealing2} that aid in this troubleshooting procedure by
enabling efficient slicing and dicing on data.  However, we have learned from
domain experts that often it is desirable to apply different models or
algorithms on the data for detailed diagnosis. Thus, most of the RAN trouble
tickets end up with experts who work directly on the raw measurement data.

\subsection{Need for Domain Specific Approach}
We now discuss the difficulties in applying machine learning for solving the RAN
performance analysis problem, thereby motivating the need for a new domain
specific solution. 

\subsubsection{Ineffectiveness of Global Modelling}
\label{sec:motivation_globalvslocal}
A common solution to applying ML on a dataset is to consider the dataset as a
single entity and build one model over the entire data. However, base stations
in a cellular network exhibit different characteristics. This renders the use of
a global model ineffective. To illustrate this problem, we conducted an
experiment where the goal was to build a model for call drops in the network. We
first run a decision tree algorithm to obtain a single model for the network.
The other extreme for this approach is to train a model per base station.
Figure~\ref{fig:spatialmodel} shows the results of this experiment which used
data collected over an 1 hour interval to ensure there is enough data for the
algorithms to produce statistically significant results. We see that the local
model is significant better, with up to 20\% more accuracy while showing much
lower variance. 

\subsubsection{Latency/Accuracy Issues with Local Models}
It is natural to think of a per base station model as the final solution to this
problem. However, this approach has issues too. Due to the difference in
characteristics of the base stations, the amount of data they collect is
different. Thus, in small intervals, they may not generate enough data to
produce valid results. This is illustrate in figure~\ref{fig:grouping} which
shows the quartiles, min and max amount of data generated and the latency
required to collect them. 

Additionally, algorithms may produce stale models with increasing latency. To
show this, we conduct an experiment with two different algorithms on data
collected over varying latencies. The first algorithm (Alg 1) builds a
classification model for connection failures, while the second (Alg 2) builds a
regression model to predict and explain throughput anomalies. The results of
this experiment is given in figure~\ref{fig:latencyissue}. The first algorithm
behavior is obvious; as it gets more data its accuracy improves due to the slow
varying nature of the underlying causes of failures. After an hour latency, it
is able to reach a respectable accuracy. However, the second algorithm's
accuracy improves initially, but falls quickly. This is counterintuitive in
normal settings, but the explanation lies in the spatio-temporal characteristics
of cellular networks. Many of the performance metrics exhibit high temporal
variability, and thus need to be analyzed in smaller intervals. In such cases,
local modeling is ineffective.

It is important to note that an obvious, but flawed, conclusion is to think that
models similar to Alg 1 would work once the data collection latency has been
incurred once. This is not true due to staleness issues which we discuss next. 

\subsubsection{Need for Model Updates}
Due to the temporal variations in cellular networks, models need to be updated
to retain their performance. To depict this, we repeated the experiment where we
built per base station decision tree model for call drops. However, instead of
training and testing on parts of the same dataset, we train on an hours worth of
data, and apply it to the next hour. Figure~\ref{fig:spatialmodel} shows that
the accuracy drops by 12\% with a stale model. Thus, it is important to keep the
model fresh by incorporating learning from the incoming data while also removing
historical learnings. Such sliding updates to ML models in a general setting is
difficult due to the overheads in retraining them from scratch. To add to this,
cellular networks consist of several thousands of base stations. This number is
on the rise with the increase in user demand and the ease of deployment of small
cells.  Thus, a per base station approach requires creating, maintaining and
updating a huge amount of models (e.g., our network consisted of over 13000 base
stations). This makes scaling hard.

\subsubsection{Why not Spatial/Spatio-Temporal Partitioning?}
The above experiments point towards the need for obtaining enough data with low
latency. The obvious solution to combating this trade-off is to
\textit{intelligently} combine data from multiple base stations. It is intuitive
to think of this as a spatial partitioning problem, since base stations in the
real world are geographically separated. Thus, a spatial partitioner which
combines data from base stations within a geographical region must be able to
give good results.  Unfortunately, this isn't the case which we motivate using a
simple example.  Consider two base stations, one situated at the center of times
square in New York and the other a mile away at a residential area. Using a
spatial partitioning scheme that divides the space into equal sized planes would
likely result in combining data from these base stations. However, this is not
desirable because of the difference in characteristics of these base stations
\footnote{In our measurements, a base station in a highly popular spot serves
    more than 300 UEs and carries multiple times uplink and downlink traffic
    compared to another base station situated just a mile from it that serves
    only 50 UEs.}. We illustrate this using the drop experiment.
Figure~\ref{fig:spatialmodel} shows the performance of a spatial model, where we
combine data from nearby base stations using a simple grid partitioner. The
results show that the spatial partitioner is not much better than the global
partitioner. We show comparisons with other smarter spatial partitioning
approaches in \ref{sec:evaluation}.

\section{\NAME{} Overview}
\label{sec:overview}
\begin{figure}[!t]
\centering
\includegraphics[width=\linewidth]{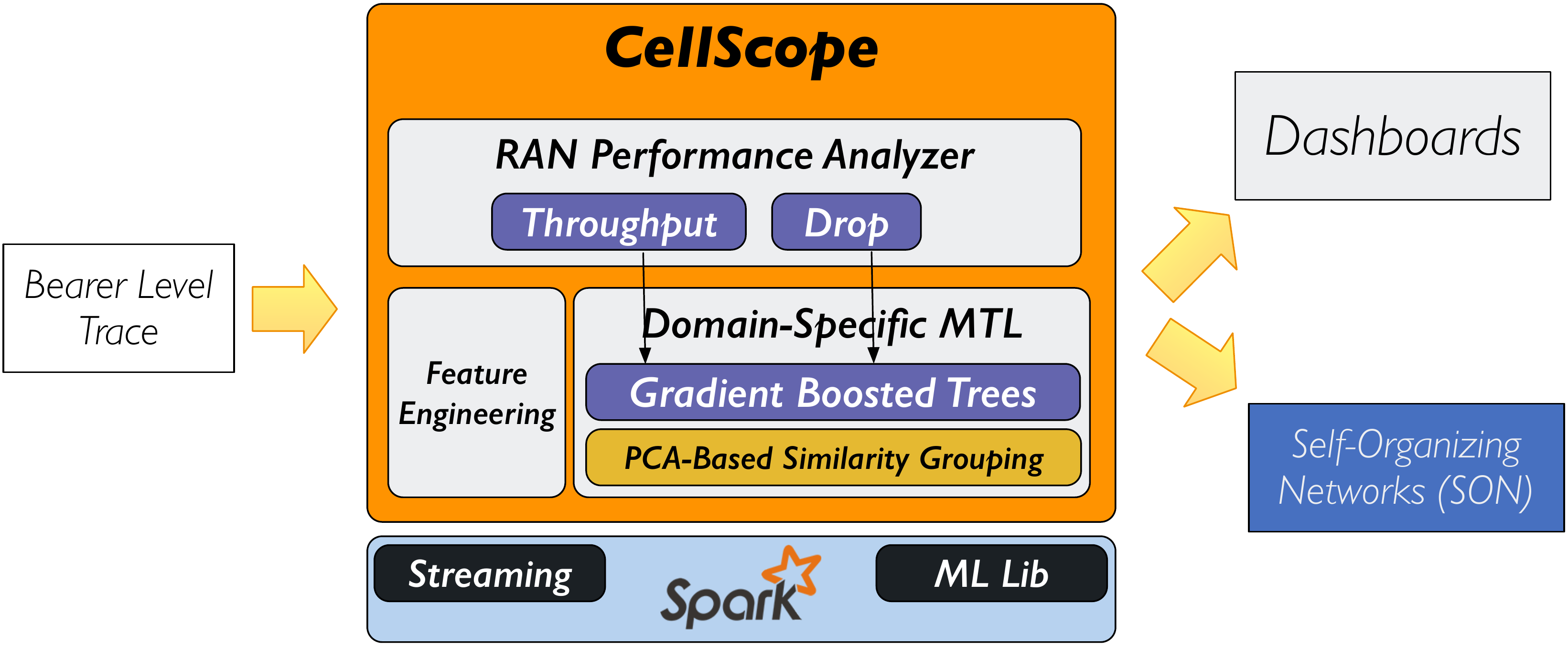}
\caption{\NAME{} System Architecture.}
\label{fig:arch} 
\end{figure}

\NAME{} presents a domain-specific formulation and application of Multi-Task
Learning (MTL) for RAN performance diagnosis.  Here, we provide a brief overview
of \NAME{} to aid the reader in following the rest of this paper.

\subsection{Problem Statement}
\NAME{}'s ultimate goal is to enable \textit{fast and accurate RAN performance
    diagnosis by resolving the trade-off between data collection latency and the
    achieved accuracy}.  The key difficulty arises from the fundamental
trade-off between \emph{having not enough data to build accurate-enough models
    in short timespans} and \emph{waiting to collect enough data that entails
    stale results} that is impossible to resolve in a general setting.
Additionally, we must support efficient modifications to the learned models to
account for the temporal nature of our setting to avoid data staleness.

\subsection{Architectural Overview}
\Cref{fig:arch} shows the high-level architecture of \NAME{}, which has the
following key components:

\noindent\textbf{Input data:} \NAME{} uses bearer-level traces that are readily
available in cellular networks (\cref{sec:lte_primer}). Base stations collect
traces independently and send them to the associated MME. 
records if required (users move, The MME merges records if required and 
hence, generate traces at multiple base stations) and uploads them to a data
center.\footnote{The transfer of traces to a data center is not fundamental.
    Extending \NAME{} to do geo-distributed learning is a future work.}

\noindent\textbf{Feature engineering:} Next, \NAME{} uses domain knowledge to
transform the raw data and constructs a set of features amenable to learning
(e.g., computing interference ratios)(\cref{sec:feature_engineering}). We also
leverage protocol details and algorithms (e.g., link adaptation) in the physical
layer.

\noindent\textbf{Domain-specific MTL:} \NAME{} uses a domain specific
formulation and application of MTL that allows it to perform accurate diagnosis
while updating models efficiently (\cref{sec:mtl}).

\noindent\textbf{Data partitioner:} To enable correct application of MTL,
\NAME{} implements a partitioner based on a similarity score derived from
Principal Component Analysis (PCA) and geographical distance
(\cref{sec:pca_partitioning}). The partitioner segregates data to be analyzed
into independent sets and produces a smaller co-located set relevant to the
analysis. This minimizes the need to shuffle data during the training process. 

\noindent\textbf{RAN performance analyzer:} This component binds everything
together to build diagnosis modules. It leverages the MTL component and uses
appropriate techniques to build call drop and throughput models. We discuss our
experience of applying these techniques to a live LTE network in
\cref{sec:findings}.

\noindent\textbf{Output:} Finally, \NAME{} can output analytics results to
external modules such as RAN performance dashboards. It can also provide inputs
to Self-Organizing Networks (SON).

\section{Resolving Latency-Accuracy Trade-off}
\label{sec:cellscope}
In this section, we present how \NAME{} uses domain-specific machine learning to
mitigate the trade-off between latency and accuracy. We first discuss a
high-level overview of RAN specific feature engineering that prepares the data
for learning (\cref{sec:feature_engineering}). Next, we describe \NAME{}'s MTL
formulation (\cref{sec:mtl}), discussing how it lets us build fast, accurate,
and incremental models. Then, we explain how \NAME{} achieves grouping that
captures commonalities among base stations using a novel PCA based partitioner
(\cref{sec:pca_partitioning}).  Finally, we summarize our approach in
\cref{sec:summary}.

\subsection{Feature Engineering} \label{sec:feature_engineering} Feature
engineering, the process of transforming the raw input data to a set of features
that can be effectively utilized by machine learning algorithms, is a
fundamental part of ML applications~\cite{FeatureSelectionSIGMOD2014}. Generally
carried out by domain experts, it is often the first step in applying learning
techniques. 

Bearer-level traces contain several hundred fields associated with LTE network
procedures. Unfortunately, many of these fields are not suitable for model
building as it is. Several fields are collected in a format that utilizes a
compact representation. For instance, block error rates need to be computed
across multiple records to account for time. Further, these records are not
self-contained, and multiple records need to be analyzed to create a feature for
a certain procedure. In \cref{sec:findings}, we describe in detail many of the
specific feature engineerings that helped \NAME{} uncover issues in the network.

\subsection{Multi-Task Learning}
\label{sec:mtl}
The latency-accuracy trade-off makes it hard to achieve both low latency and
high accuracy in applied machine learning tasks (\cref{sec:motivation}). The
ideal-case scenario in \NAME{} is if infinite amount of data is available per
base station with zero latency. In this scenario, we would have a learning task
for each base station that produce a model as an output with the best achievable
accuracy. In reality, our setting has several tasks, each with its own data.
However, each task does not have enough data to produce models with acceptable
accuracy in a given latency budget. This makes our setting an ideal candidate
for multi-task learning (MTL), a cutting-edge research area in machine learning.
The key idea behind MTL is to \emph{learn from other tasks by weakly coupling
    their parameters so that the statistical efficiency of many tasks can be
    boosted}~\cite{MTLICML1993,MTLNIPS1996,RMLKDD2004,InductiveBiasLearning2000}.
Specifically, if we are interested in building a model of the form 
\begin{equation}\label{eqn:mlmodel}
 h(x) = m (f_1(x), f_2(x), ... , f_k(x)) 
\end{equation}
where $m$ is a model composed of features $f_1$ through $f_k$, then the
traditional MTL formulation, given dataset $\mathcal{D }= \{(x_i, y_i, bs_i):
i=1,...,n\}$, where $x_i \in \mathbb{R}^d, y_i \in \mathbb{R}$ and $bs_i$
denotes the $i^{th}$ base station, is to learn \begin{equation}\label{eqn:mtl}
    h(x) = m_{bs} (f_1(x), f_2(x), ... , f_k(x)) \end{equation} where $m_{bs}$
is a per base station model. 

In this MTL formulation, the core assumption is a shared structure or dependency
across each of the learning problems. Unfortunately, in our setting, the base
stations do not share a structure at a global level (\cref{sec:motivation}). Due
to their geographic separation and the complexities of wireless signal
propagation, the base stations share a spatio-temporal structure instead. Thus,
we proposes a new domain-specific MTL formulation.

\subsubsection{\NAME{}'s MTL Formulation}
In order to address the difficulty in applying MTL due to the violation of task
dependency assumption in RANs, we can leverage domain-specific characteristics.
Although independent learning tasks (learning per base station) are not
correlated with each other, they exhibit specific non-random structure.  For
example, the performance characteristics of base stations nearby are influenced
by similar underlying features. Thus, we propose exploiting this knowledge to
segregate learning tasks into groups of dependent tasks on which MTL can be
applied. MTL in the face of dependency violation has been studied in the machine
learning literature in the recent past~\cite{TreeMTLICML2010, RobustMTLKDD2012}.
However, they assume that each group has its own set of features. This is not
entirely true in our setting, where multiple groups may share most or all
features but still need to be treated as separate groups.  Furthermore, some of
the techniques used for automatic grouping without a priori knowledge  are
computationally intensive. 

Assuming we can club  learning tasks into groups, we can rewrite the MTL eq.
\eqref{eqn:mtl} to captures this structure as
\begin{equation}\label{eqn:mtl_cellscope}
h(x) = m_{g(bs)} (f_1(x), f_2(x), ... , f_k(x))
\end{equation}
where $m_{g(bs)}$ is the per-base station model in group $g$. We describe a
simple technique to achieve this grouping based on domain knowledge in
\cref{sec:pca_partitioning} and experimentally show that just grouping by itself
can achieve significant gains in \cref{sec:evaluation}.

In theory, the MTL formulation in eq. \eqref{eqn:mtl_cellscope} should suffice
for our purposes as it would perform much better by capturing the inter-task
dependencies using grouping. However, this formulation still builds an
independent model for each base station.  Building and managing a large amount
of models leads to significant performance overhead and would impede our goal of
scalability. Scalable application of MTL in a general setting is an active area
of research in machine learning~\cite{ScalableMTL}, so we turn to
problem-specific optimizations to address this challenge.

The model $m_{g(bs)}$ could be built using any class of learning functions. In
this paper, we restrict ourselves to functions of the form $F(x) = w.x$ where
$w$ is the weight vector associated with a set of features $x$. This simple
class of function gives us tremendous leverage in using standard algorithms that
can easily be applied in a distributed setting, thus addressing the scalability
issue. In addition to scalable model building, we must also be able to update
the built models fast. However, machine learning models are typically hard to
update in real time. To address this challenge, we discuss a hybrid approach to
building the models in our MTL setting next.

\subsubsection{Hybrid Modeling for Fast Model Updates}
Estimation of the model in eq. \eqref{eqn:mtl_cellscope} could be posed as an
$\ell_1$ regularized loss minimization problem~\cite{Lasso1994}:
\begin{equation}\label{eqn:lasso}
    min \sum L(h(x:f_{bs}), y) + \lambda ||R(x:f_{bs})||
\end{equation}
where $L(h(x:f_{bs}), y)$ is a non-negative loss function composed of parameters
for a particular base station, hence capturing the error in the prediction for
it in the group, and $\lambda > 0$ is a regularization parameter scaling the
penalty $R(x:f_{bs})$ for the base station.  However, the temporal and streaming
nature of the data collected means that the model must be refined frequently for
minimizing staleness. 

Fortunately, grouping provides us an opportunity to solve this.  Since the base
stations are grouped into correlated task clusters, we can decompose the
features used for each base station into a \textit{shared common set} $f_c$ and
a \textit{base station specific} set $f_s$. Thus, we can modify the eq.
\eqref{eqn:lasso} as minimizing
\begin{equation}\label{eqn:lasso_mtl}
    \sum\Bigg(\sum L(h(x:f_s), y) + \lambda ||R(x:f_s)||\Bigg) +  \lambda
    ||R(x:f_c)||
\end{equation}
where the inner summation is over dataset specific to each base station. This
separation gives us a powerful advantage. Since we already grouped base
stations, the feature set $f_s$ is minimal, and in most cases just a weight
vector on the common feature set rather than a complete new set of features. 

Because the core common features do not change often, we need to update only the
base station-specific parts in the model frequently, while the common set can be
reused. Thus, we end up with a hybrid offline-online model. Furthermore, the
choice of our learning functions lets us apply stochastic
methods~\cite{L1SML2011} which can be efficiently parallelized.

\subsubsection{Anomaly Detection Using Concept Drift}
A common use case of learning tasks for RAN performance analysis is in detecting
anomalies. For instance, an operator may be interested in learning if there is a
sudden increase in call drops. At the simplest level, it is easy to answer this
question by simply monitoring the number of call drops at each base station.
However, just a yes or no answer to such questions are seldom useful. If there
is a sudden increase in drops, then it is useful to understand if the issue
affects a complete region and the root cause of it.

Our MTL approach and the ability to do fast incremental learning enables a
better solution for anomaly detection and diagnosis.  Concept drift is a term
used to refer the phenomenon where the underlying distribution of the training
data for a machine learning model changes~\cite{ConceptDriftSurvey}. \NAME{}
leverages this to detect anomalies as concept drifts and proposes a simple
technique for it. Since we process incoming data in mini-batches
(\cref{sec:implementation}), each batch can be tested quickly on the existing
model for significant accuracy drops. An anomaly occurring just at a single base
station would be detected by one model, while one affecting a larger area would
be detected by many. Once anomaly has been detected, finding the root cause is
as easy as updating the model and comparing it with the old one.

\subsection{Data Grouping for MTL} \label{sec:pca_partitioning} Having discussed
\NAME{}'s MTL formulation, we now turn our focus towards how \NAME{} achieves
efficient grouping of cellular datasets that enables accurate learning.  Our
data partitioning is based on Principal Component Analysis (PCA), a widely used
technique in multivariate analysis~\cite{PCA}. PCA uses an orthogonal coordinate
transformation to map a given set of points into a new coordinate space. Each of
the new subspaces are commonly referred to as a principal component. Since the
coordinate space is equal to or smaller than the original , PCA is used for
dimensionality reduction. 

In their pioneering work, Lakhina et.al.~\cite{AnomaliesSigcomm2004} showed the
usefulness of PCA for network anomaly detection. They observed that it is
possible to segregate normal behavior and abnormal (anomalous) behavior using
PCA---the principal components explain most of the normal behavior while the
anomalies are captured by the remaining subspaces. Thus, by filtering normal
behavior, it is possible to find anomalies that may otherwise be undetected.

While the most common usecase for PCA has been dimensionality reduction (in
machine learning domains) or anomaly detection (in networking domain), we use it
in a novel way, to enable grouping of datasets for multi-task learning. Due to
the lack of the ability to collect sufficient amount of data from individual
base stations, detecting anomalies in them will not yield results. However, the
data would still yield an explanation of normal behavior. We use this
observation to partition the dataset. We describe our notation first.

\subsubsection{Notation}
Since bearer level traces are collected continuously, we consider a buffer of
bearers as a \textit{measurement matrix} $A$. Thus, $A$ consists of $m$ bearer
records, each having $n$ observed parameters making it an $m \times n$
time-series matrix. It is to be noted that $n$ is in the order of a few 100
fields, while $m$ can be much higher depending on how long the buffering
interval is. We enforce $n$ to be fixed in our setting---every measurement
matrix must contain $n$ columns. To make this matrix amenable to PCA analysis,
we adjust the columns to have zero mean. By applying PCA to any measurement
matrix $A$, we can obtain a set of $k$ principal components ordered by amount of
data variance they capture.  

\subsubsection{PCA Similarity}
It is intuitive to see that many measurement matrices may be formed based on
different criteria. Suppose we are interested in finding if two measurement
matrices are \textit{similar}. One way to achieve this is to compare the
principal components of the two matrices. Krzanowski~\cite{PCASimilarity}
describes such a \textit{Similarity Factor} ($SF$). Consider two matrices $A$
and $B$ having the same number of columns, but not rows. The similarity factor
between $A$ and $B$ is defined as:     
\[ SF = trace(LM'ML') = \sum_{i=1}^{k}\sum_{j=1}^{k}\cos^2\theta_{ij}\]
where $L$, $M$ are the first $k$ principal components of $A$ and $B$, and
$\theta_{ij}$ is the angle between the $i^{th}$ component of $A$ and the
$j^{th}$ component of $B$. Thus, similarity factor considers all combinations of
$k$ components from both the matrices. 

\subsubsection{\NAME{}'s Similarity Metric}
Similarity in our setting bears a slightly different notion: we do not want
strict similarity between measurement matrices, but only need similarity between
corresponding principal components. This ensures that algorithms will still
capture the underlying major influences and trends in observation sets that are
not exactly similar. Unfortunately, $SF$ does not fit our requirements; hence,
we propose a simpler metric.

Consider two measurement matrices $A$ and $B$ as before, where $A$ is of size
$m_{A} \times n$ and $B$ is of size $m_{B} \times n$. By applying PCA on the
matrices, we can obtain $k$ principal components using a heuristic. We obtain
the first $k$ components which capture $95\%$ of the variance. From the PCA, we
obtain the resulting weight vector, or \textit{loading}, which is a $n \times k$
matrix: for each principal component in $k$, the loading describes the weight on
the original $n$ features.  Intuitively, this can be seen as a rough measure of
the influence of each of the $n$ features on the principal components. The
Euclidean distance between the corresponding loading matrices gives us
similarity:
\[ SF_{\NAME{}} = \sum_{i=1}^{k}d(a_i, b_i) =
    \sum_{i=1}^{k}\sum_{j=1}^{n}|a_{ij}-b_{ij}|\]
where $a$ and $b$ are the column vectors representing the loadings for the
corresponding principal components from $A$ and $B$. Thus, $SF_{\NAME{}}$
captures how closely the underlying features explain the variation in the data.

Due to the complex interactions between network components and the wireless
medium, many of the performance issues in RANs are geographically tied (e.g.,
congestion might happen in nearby areas, and drops might be
concentrated)\footnote{Proposals for conducting geographically weighted PCA
    (GW-PCA) exist~\cite{GWPCA}, but they are not applicable since they assume a
    smooth decaying user provided bandwidth function.}. However, $SF_{\NAME{}}$
doesn't capture this phenomenon because it only considers similarity in normal
behavior. Consequently, it is possible for anomaly detection algorithms to miss
geographically-relevant anomalies. To account for this domain-specific
characteristic, we augment our similarity metric to also capture the
geographical closeness by weighing the metric by geographical distance between
the two measurement matrices. Our final similarity metric is\footnote{A
    similarity measure for multivariate time series is proposed
    in~\cite{pcasimilaritytimevariate}, but it is not applicable due to its
    stricter form and dependence on finding the right eigenvector matrices to
    extend the Frobenius norm.}:
\[ SF_{\NAME{}} = w_{distance_{(A,B)}} \times \sum_{i=1}^{k}\sum_{j=1}^{n}|a_{ij}-b_{ij}|\]

\subsubsection{Using Similarity Metric for Partitioning} With similarity metric,
\NAME{} can now partition bearer records. We first group the bearers into
measurement matrices by segregating them based on the cell on which the bearer
originated. The grouping is based on our observation that the cell is the lowest
level at which an anomaly would manifest. We then create a graph $G(V,E)$ where
the vertices are the individual cell measurement matrices.  An edge is drawn
between two matrices if the $SF_{\NAME{}}$ between them is below a threshold. To
compute $SF_{\NAME{}}$, we simply use the geographical distance between the
cells as the weight. Once the graph has been created, we run connected
components on this graph to obtain the partitions. The use of connected
component algorithm is not fundamental, it is also possible to use a clustering
algorithm instead. For instance, a k-means clustering algorithm that could
leverage $SF_{\NAME{}}$ to merge clusters would yield similar results.

\subsubsection{Managing Partitions Over Time} One important consideration is
managing and handling group changes over time.  To detect group changes, it is
necessary to establish correspondence between groups across time intervals. Once
this correspondence is established, \NAME{}'s hybrid modeling makes it easy to
accommodate changes. Due to the segregation of our model into common and base
station specific components, small changes to the group do not affect the common
model. In these cases, we can simply bootstrap the new base station using the
common model, and then start learning specific features. On the other hand, if
there are significant changes to a group, then the common model may no longer be
valid, which is easy to detect using concept drift. In such cases, the offline
model could be rebuilt. 

\subsection{Summary}
\label{sec:summary}
We now summarize how \NAME{} resolves the fundamental trade-off between latency
and accuracy. To cope with the fact that individual base stations cannot produce
enough data for learning in a given time budget, \NAME{} uses MTL.  However, our
datasets violate the assumption of learning task dependencies. As a solution, we
proposed a novel way of using PCA to group data into sets with the same
underlying performance characteristics. Directly applying MTL on these groups
would still be problematic in our setting due to the inefficiencies with model
updates. To solve this, we proposed a new formulation for MTL which divides the
model into an offline and online hybrid.  On this formulation, we proposed using
simple learning functions are amenable to incremental and distributed execution.
Finally, \NAME{} uses a simple concept drift detection to find and diagnose
anomalies.

\section{Implementation}
\label{sec:implementation}

We have implemented \NAME{} on top of Spark~\cite{RDD}, a big data cluster
computing framework. In this section, we describe its API that exposes our
commonality based grouping based on PCA (\cref{sec:api}), and implementation
details on the hybrid offline-online MTL models (\cref{sec:hybrid_modeling}). 

\subsection{Data Grouping API}
\label{sec:api}
\NAME{}'s grouping API is built on Spark Streaming~\cite{SparkStreamingSOSP13},
since the data arrives continuously, and we need to operate on this data in a
streaming fashion. Spark Streaming already provides support for windowing
functions on streams of data, thus we extended the windowing functionality with
three APIs in listing \ref{lst:api}. In this section, we use the words grouping
and partitions interchangeably.

\begin{figure}[t]
\begin{lstlisting}[language=scala, label=lst:api,
  caption={\NAME{}'s Grouping API}]
grouped = DStream.groupBySimilarityAndWindow
        (windowDuration, slideDuration)
reduced = DStream.reduceBySimilarityAndWindow
        (func, windowDuration, slideDuration)
 joined = DStream.joinBySimilarityAndWindow
        (windowDuration, slideDuration)
\end{lstlisting}
\end{figure}

Both the APIs leverage the \texttt{DStream} abstraction provided by Spark
Streaming. The \texttt{groupBySimilarityAndWindow} takes the buffered data from
the last window duration, applies the similarity metric to produce outputs of
grouped datasets (multiple DStreams) every slide duration. The
\texttt{reduce\-By\-Similari\-t\-y\-And\-Window} allows an additional user
defined associative reduction operation on the grouped datasets. Finally, it
also provides a \texttt{joinBySimilarityAndWindow} which joins multiple streams
using similarity. We found these APIs sufficient for most of the operations,
including group changes. 

\subsection{Hybrid MTL Modeling}
\label{sec:hybrid_modeling}
We use Spark's machine learning library, MLlib~\cite{MLI13} for implementing our
hybrid MTL model. MLlib contains the implementation of many distributed learning
algorithms. To leverage the many pre-existing algorithms in Mllib, we
implemented our multi-task learning hybrid model as an ensemble
method~\cite{ensemble}. By definition, ensemble methods use multiple learning
algorithms to obtain better performance. Given such methods, it is easy to
implement our hybrid online-offline model; the shared features can be
incorporated as a static model and the per base station model can be a separate
input. 

We modified the MLlib implementation of Gradient Boosted Tree (GBT)~\cite{GBT}
model, an ensemble of decision trees. This implementation supports both
classification and regression, and internally utilizes stochastic methods. Our
modification supports a cached offline model in addition to online models. To
incorporate incremental and window methods, we simply add more models to the
ensemble when new data comes in. This is possible due to the use of stochastic
methods. We also support weighing the outcome of the ensemble, so as to give
more weights to the latest models. 

\section{Evaluation}
\label{sec:evaluation}

\begin{figure*}[t!]
  \centering
  \begin{subfigure}{.33\textwidth}
  \centering
  \includegraphics[width=\linewidth]{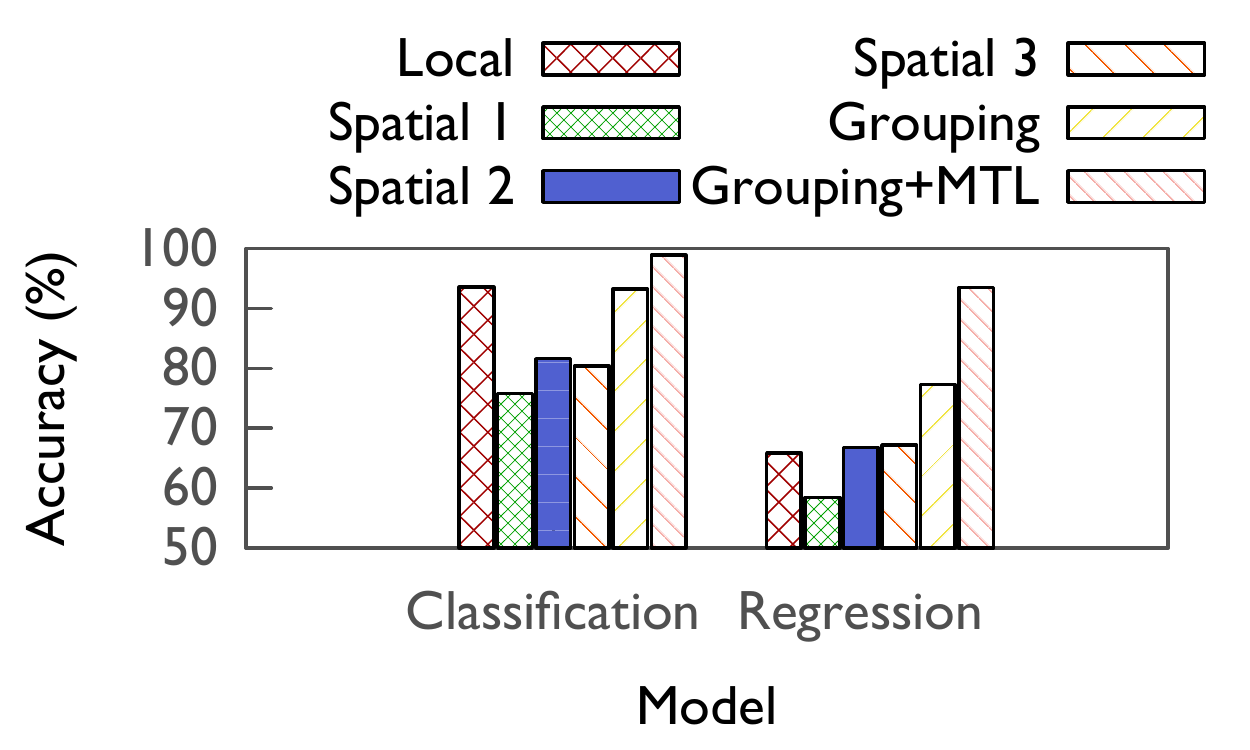}
  \captionof{figure}{\NAME{}'s partitioning by itself is able to provide significant
      gains. MTL provides further gains}
  \label{fig:overall}
  \end{subfigure}%
  \begin{subfigure}{.33\textwidth}
    \centering
    \includegraphics[width=\linewidth]{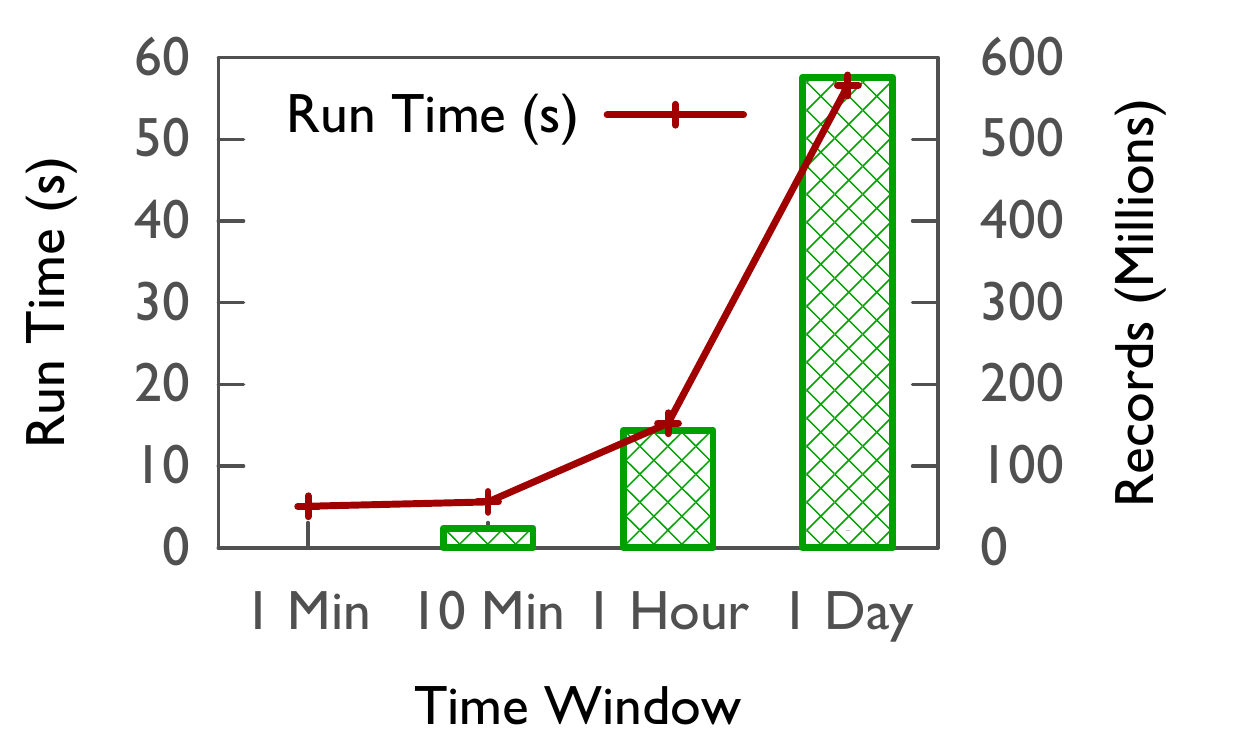}
    \captionof{figure}{Partitioning
        overheads are minimal}
  \label{fig:scaling}
  \end{subfigure}
  \begin{subfigure}{.33\textwidth}
  \centering
  \includegraphics[width=\linewidth]{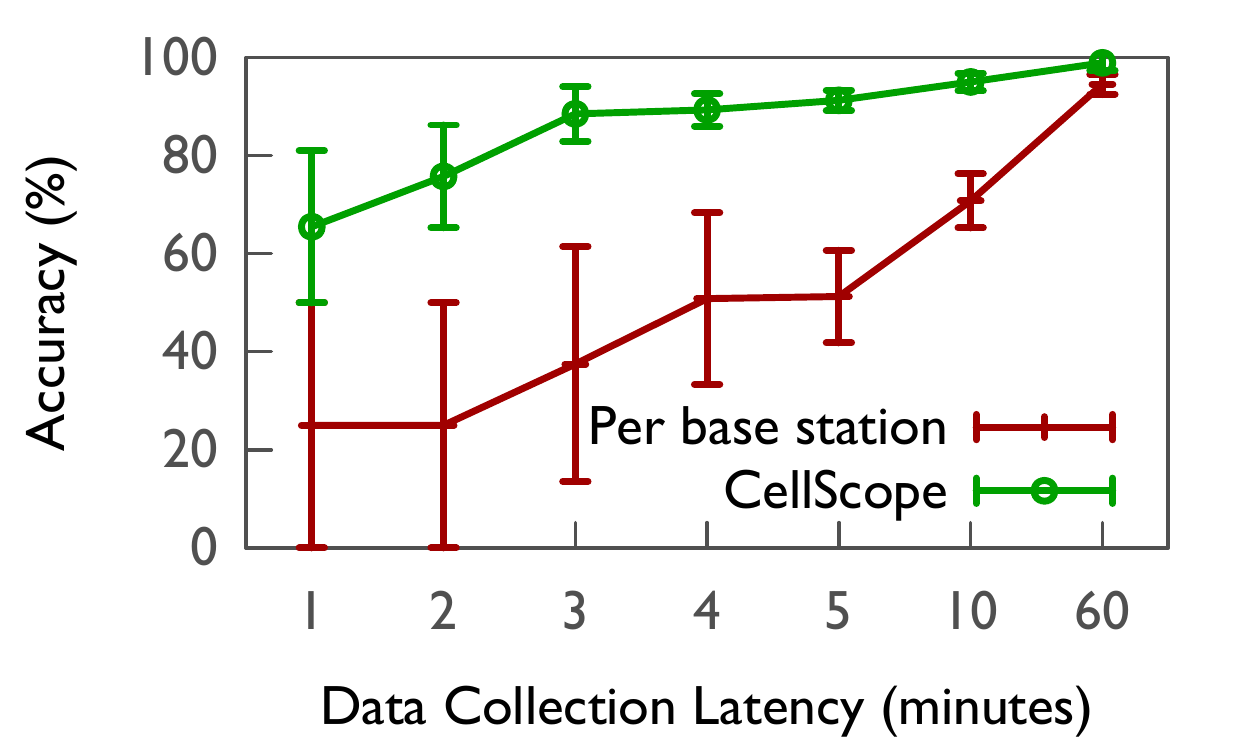}
  \captionof{figure}{\NAME{} achieves up to 2.5$\times$ accuracy improvements in
      drop rate classification}
  \label{fig:classification}
 \end{subfigure}
   \centering
   \begin{subfigure}{.33\textwidth}
  \centering
  \includegraphics[width=\linewidth]{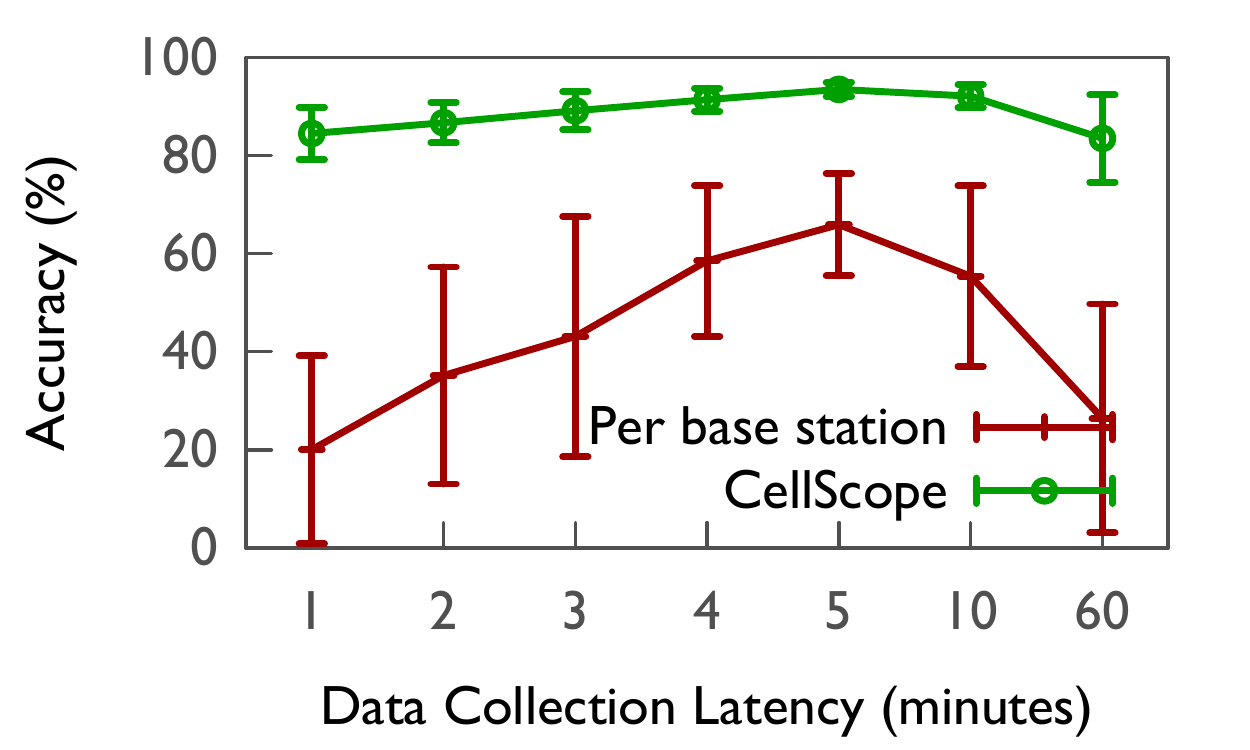}
  \captionof{figure}{Improvements in throughput model regression go up to
      4.4$\times$}
  \label{fig:regression}
  \end{subfigure}%
    \begin{subfigure}{.33\textwidth}
    \centering
    \includegraphics[width=\linewidth]{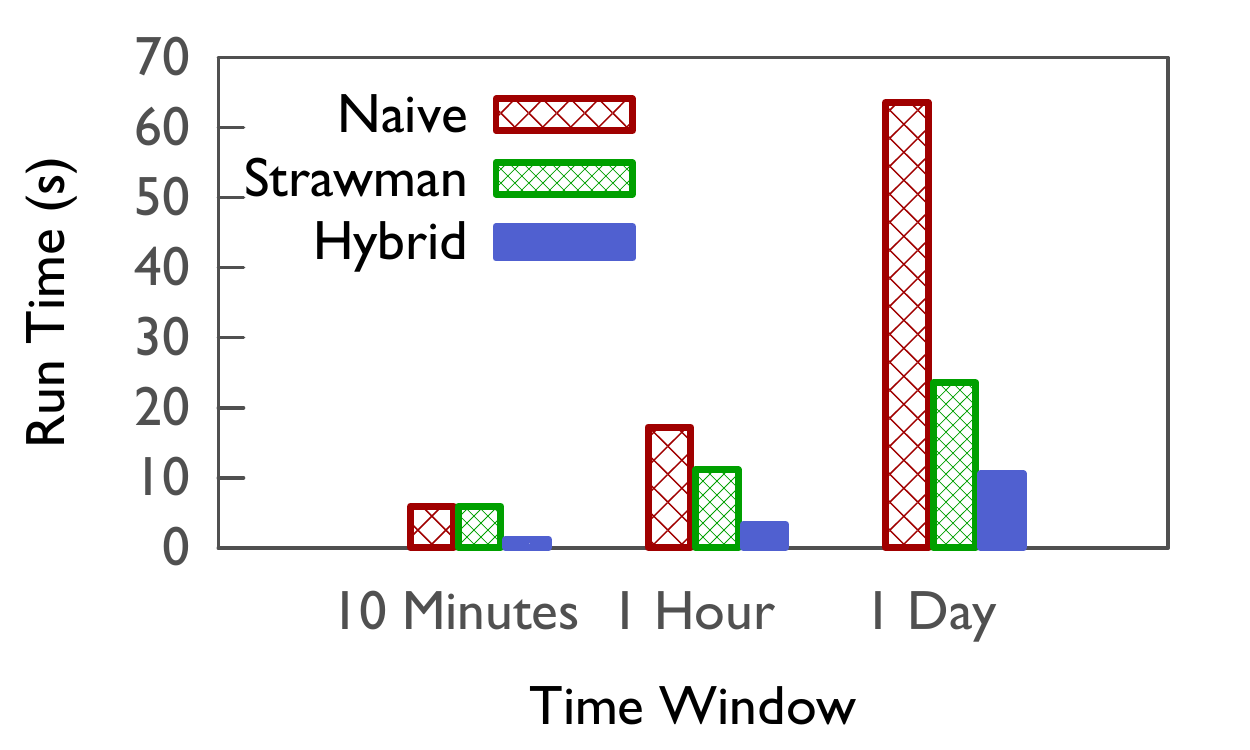}
    \captionof{figure}{Hyrid model reduces update time by up to
        4.8$\times$}
  \label{fig:hybrid}
  \end{subfigure}
  \begin{subfigure}{.33\textwidth}
  \centering
  \includegraphics[width=\linewidth]{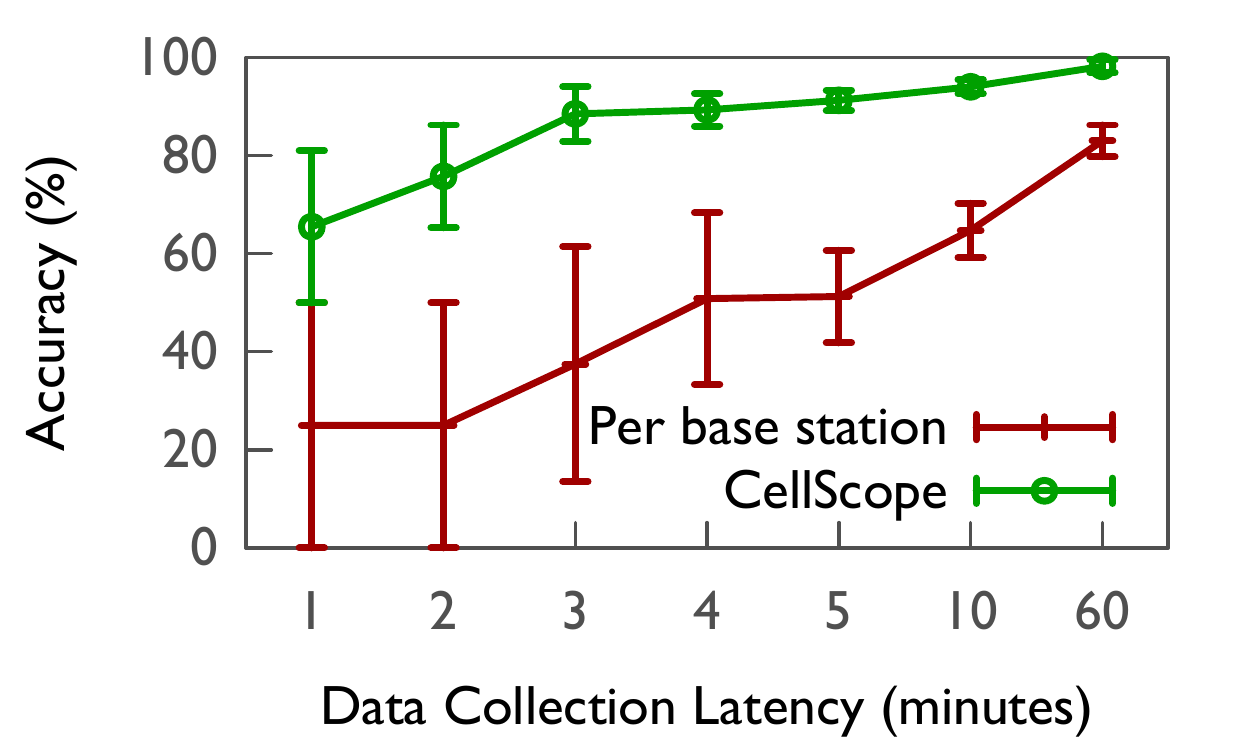}
  \captionof{figure}{Online training due to the hybrid model helps incur
      almost no loss in accuracy due to staleness.}
  \label{fig:staleness}
 \end{subfigure}
 \caption{\NAME{} is able to achieve high accuracy while reducing the data
     collection latency.}
\end{figure*}

We have evaluated \NAME{} through a series of experiments on real-world cellular
traces from a live LTE network from a large geographical area. Our results are
summarized below: 
\begin{itemize}
    \item \NAME{}'s similarity based grouping provides up to 10\% improvement in
        accuracy on its own compared to the best case scenario of space
        partitioning schemes.  
    \item With MTL, \NAME{}'s accuracy improvements
        range from 2.5$\times$ to
        4.4$\times$ over different collection latencies.  
    \item Our hybrid online-offline model is able to reduce model update times 
        upto 4.8$\times$ and is able to learn changes in an online fashion with
        virtually no loss in accuracy.
\end{itemize}
We discuss these results in detail in the rest of this section.
\\
\textbf{Evaluation Setup:} Due to the sensitive nature of our dataset, our
evaluation environment is a private cluster consists of 20 machines. Each
machine consists of 4 CPUs, 32GB of memory and a 200GB magnetic hard disk.    
 \\
 \textbf{Dataset:} We collected data from a major metro-area LTE network
 occupying a large geographical area for a time period of over 10 months. The
 network serves more than 2 million active users and carries over 6TB of traffic
 per hour. 

 \subsection{Benefits of Similarity Based Grouping}
 We first attempt to answer the question \textit{"How much benefits do the
     similarity based grouping provide?"}. To answer this question, we conducted
 two experiments, each with a different learning algorithm. The first
 experiment, whose aim is to detect call drops, uses a classification algorithm
 while the second, whose aim is to predict throughput, uses a regression
 algorithm. We chose these to evaluate the benefits in two different classes of
 algorithms.  In both these cases, we pick the data collection latency where the
 per base station model gives the best accuracy, which was 1 hour for
 classification and 5 minutes for regression. In order to compare the benefits
 of our grouping scheme alone, we build a single model per group instead of
 applying MTL. We compare the accuracy obtained with three different space
 partitioning schemes. The first scheme (Spatial 1) just partitions space into
 grids of equal size. The second (Spatial 2) uses a sophisticated space-filling
 curve based approach~\cite{CellIQNSDI15} that could create dynamically size
 partitions. Finally, the third (Spatial 3) creates partitions using base
 stations that are under the same cellular region. The results are shown in
 \cref{fig:overall}.

\NAME{}'s similarity grouping performs as good as the per base station model
which gives the highest accuracy. It is interesting to note the performance of
spatial partitioning schemes which ranges from 75\% to 80\%. None of the spatial
schemes come close to the similarity grouping results. This is because the drops
are few, and concentrated. Spatial schemes club base stations not based on
underlying drop characteristics, but only based on spatial proximity. This
causes the algorithms to underfit or overfit. Since our similarity based
partitioner groups base stations using the drop characteristics, it is able to
do as much as 17\% better than the spatial schemes.

The benefits are even higher in the regression case. Here, the per base station
model is unable to get enough data to build an accurate model and hence is only
able to achieve around 66\% accuracy. Spatial schemes are able to do slightly
better than that. Our similarity based grouping emerges as a clear winner in
this case with 77.3\% accuracy. This result depicts the highly variable
performance characteristics of the base stations, and the need to capture them
for accuracy.

These benefits do not come at the cost of higher computational overhead to do
the grouping. \Cref{fig:scaling} shows the overhead of performing similarity
based grouping on various dataset sizes. It is clear that even very large
datasets can be easily partitioned with very little overhead.

\subsection{Benefits of MTL}
Next, we characterize the benefits of \NAME{}'s use of MTL. To do this, we
repeated the experiment before, and apply MTL to the grouped data to see if the
accuracy improves compared to the earlier approach of a single model per group.
The results are presented in \cref{fig:overall}. The ability of MTL to learn and
improve models from other similar base stations' data results in an increase in
the accuracy. Over the benefits of grouping, we see an improvement of 6\% in the
connection drop diagnosis experiment, and 16.2\% in the case of throughput
prediction experiment. The higher benefits in the latter comes from \NAME{}'s
ability to capture individual characteristics of the base station. This ability
is not so crucial in the former because of the limited variation in individual
characteristics over those found by the grouping. 
 
\subsection{Combined Benefits of Grouping and MTL}
We now evaluate the combined benefits of grouping and MTL under different data
collection latencies. Here, we are interested in evaluating how \NAME{} handles
the latency accuracy trade-off. To do this, we do the same classification and
regression experiments, but on different data collection latencies instead of
one. We show the results from the classification experiment in
\cref{fig:classification} and that from the regression experiment in
\cref{fig:regression}, which compares \NAME{}'s accuracy against a per base
station model's.

When the opportunity to collect data at individual base stations is limited,
\NAME{} is able to leverage our MTL formulation to combine data from multiple
base stations, and build customized models to improve the accuracy. The benefits
of \NAME{} ranges up to 2.5$\times$ in the classification experiment, to
4.4$\times$ in the regression experiment. Lower latencies are problematic in the
  classification experiment due to the extremely low probability of drops, while
  higher latencies are a problem in the regression experiment due to the
  temporal changes in performance characteristics.

\subsection{Hybrid model benefits}
Finally, we evaluate the benefits of our hybrid modeling. Here, we are
interested in learning how much overhead does it reduce during model updates,
and can it do online learning. 

To answer the first question, we conducted the following experiment: we
considered three different data collection latencies: 10 minute, 1 hour and 1
day. We then learn a decision tree model on this data in a tumbling window
fashion. So for the 10 minute latency, we collect data for 10 minutes, then
build a model, wait another 10 minutes to refine the model and so on. We
compare our hybrid model strategy to two different strategies: a naive approach
which rebuilds the model from scratch every time, and a better, strawman
approach which reuses the last model, and makes changes to it. Both builds a
single model while \NAME{} uses our hybrid MTL model and only updates the
online part of the model. The results of this experiment is shown in
\cref{fig:hybrid}.

The naive approach incurs the highest overhead, which is obvious due to the
need to rebuild the entire model from scratch. The overhead increases with the
increase in input data. The strawman approach, on the other hand, is able
to avoid this heavy overhead. However, it still incurs overheads with larger
input because of its use of a single model which requires changes to many
parts of the tree. \NAME{} incurs the least overhead, due to its use of
multiple models. When data accumulates, it only needs to update a
part of an existing tree, or build a new tree. This strategy results in a
reduction of up to 2.2$\times$ to 4.8$\times$ in model building time for \NAME{}.

To wrap up, we evaluated the performance of the hybrid strategy on different
data collection intervals. Here we are interested in seeing if the hybrid model
is able to adapt to data changes and provide reasonable accuracies. We use the
connection drop experiment again, but do it in a different way. At different
collection latencies, we build the model at the beginning of the collection and
use the model for the next interval. Hence, for the 1 minute latency, we build a
model using the first minute data, and use the model for the second minute
(until the whole second minute has arrived). The results are shown in
\cref{fig:staleness}. We see here that the per base station model suffers an
accuracy loss at higher latencies due to staleness, while \NAME{} incurs almost
zero loss in accuracy. This is because it doesn't wait until the end of the
interval, and is able to incorporate data in real time.

\section{RAN Performance Analysis Using \NAME{}}
\label{sec:findings}

\begin{table}
    \footnotesize
\begin{tabular}{|c|p{6.5cm}|}
\hline
\multicolumn{2}{|c|}{\bf LTE Physical Layer Parameters} \\
\hline
{\it Name} & {\centering\hspace*{2cm} {\it Description}}\\
\hline
RSRP & Reference Signal Received Power: Average of reference singal power (in
watts) across a specified bandwidth. Used for cell selection and handoff. \\
RSRQ & Reference Signal Recieved Quality: Indicator of interference experienced by the UE. Derived from RSRP and interference metrics.\\
CQI & Channel Quality Indicator:  Carries information on how good/bad communication channel quality is.\\
SINR & Signal to Interference plus Noise Ratio: Indicates the ratio of the power of the signal to the interference power and background noise.\\
BLER & Block Error Ratio/Rate: Ratio of the number of erroneous blocks received to the total blocks sent.\\
PRB & Physical Resource Block: The specific number of subcarriers allocated for
a predetermined amount of time for a user.\\
\hline
\end{tabular}
\caption{A description of key parameters in LTE physical layer}
\label{tbl:phyparams}
\end{table}

To validate our system in real world, we now show how domain experts can use \NAME{} to build
efficient RAN performance analysis solutions. To analyze RAN performance, we consider two metrics that are of significant
importance for end-user experience: \textbf{throughput} and
\textbf{connection drops}. Our findings from the analysis are summarized below: 
\begin{itemize}
    \item Our bearer performance analysis reveals interesting findings on
the inefficiencies of P-CQI detection mechanism and link adaptation
algorithm. Both of these findings were previously unknown to the operator. 
\item We find that connection drop is mainly due to uplink SINR and
    then downlink SINR, and that RSRQ is more reliable than downlink CQI. 
\item Our cell performance analysis shows that many unknown connection drops can be explained
    by coverage and uplink interference, and that throughput is seriously impacted by
    inefficient link adaptation algorithm.
\end{itemize}

\subsection{Analyzing Call Drop Performance}
\label{sec:retain}

\begin{figure*}[!htb]
\centering
\begin{subfigure}{.24\textwidth}
  \centering
  \includegraphics[width=\linewidth]{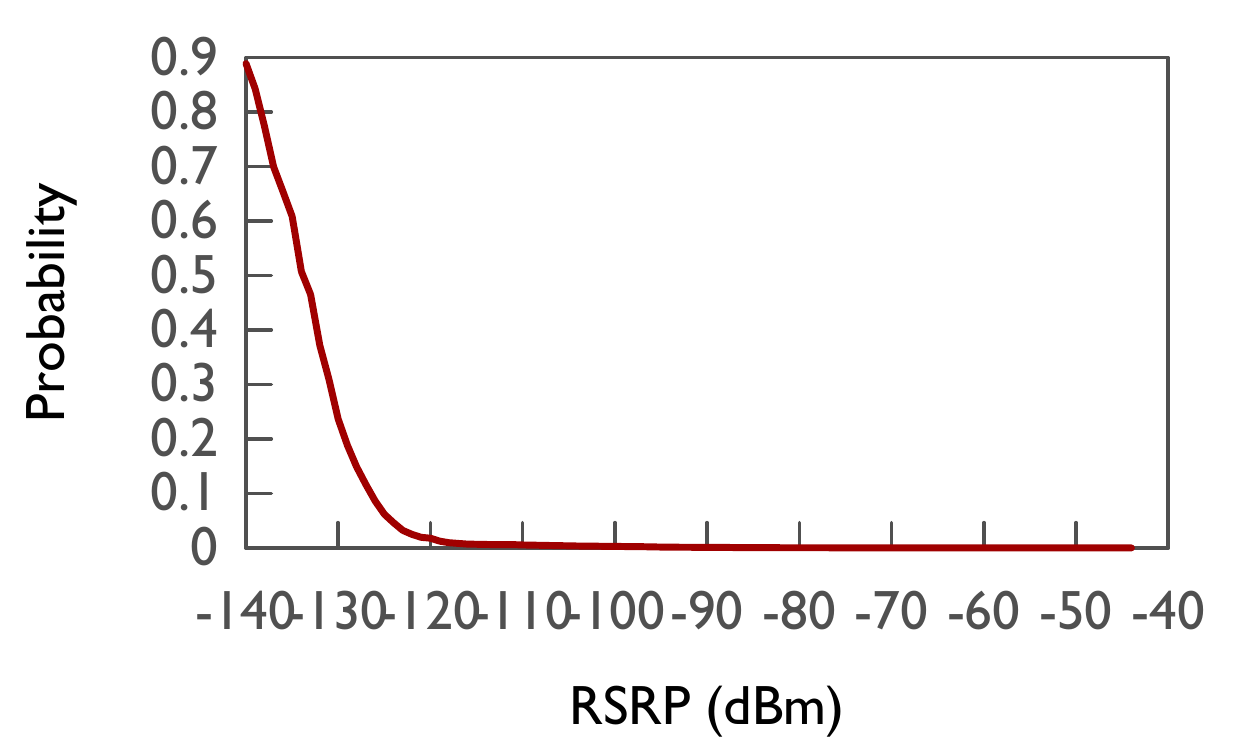}
  \captionof{figure}{Coverage}
  \label{fig:coverage_cdf}
  \end{subfigure}%
  \begin{subfigure}{.24\textwidth}
  \centering
  \includegraphics[width=\linewidth]{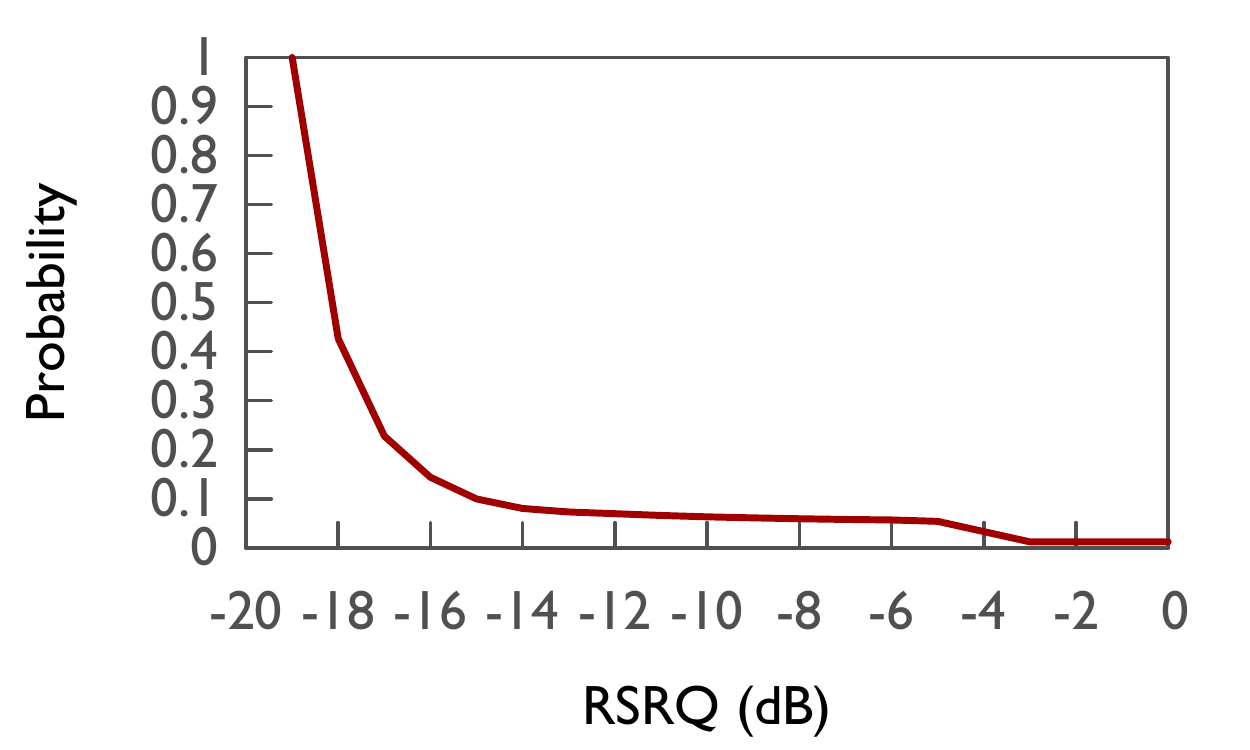}
  \captionof{figure}{Downlink RSRQ}
  \label{fig:interference_cdf}
  \end{subfigure}
  \begin{subfigure}{.24\textwidth}
  \centering
  \includegraphics[width=\linewidth]{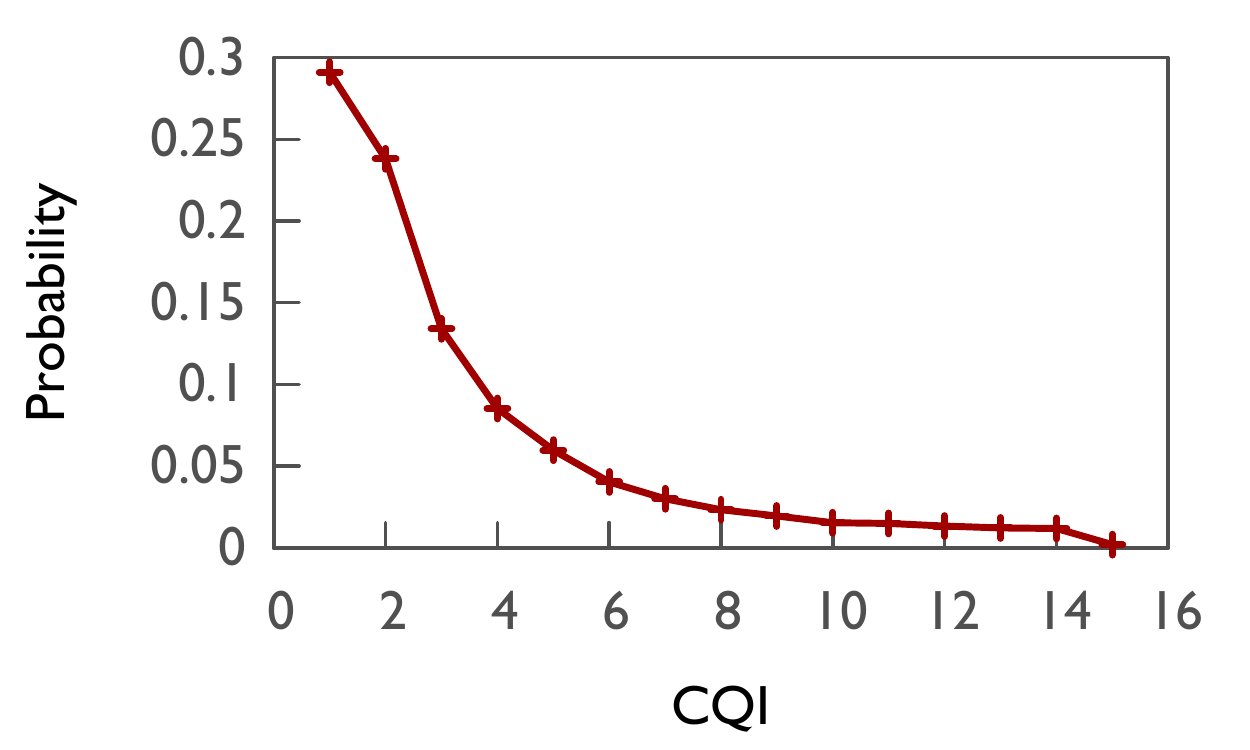}
  \captionof{figure}{Downlink CQI}
  \label{fig:cqi}
 \end{subfigure}
 \begin{subfigure}{.24\textwidth}
  \centering
  \includegraphics[width=\linewidth]{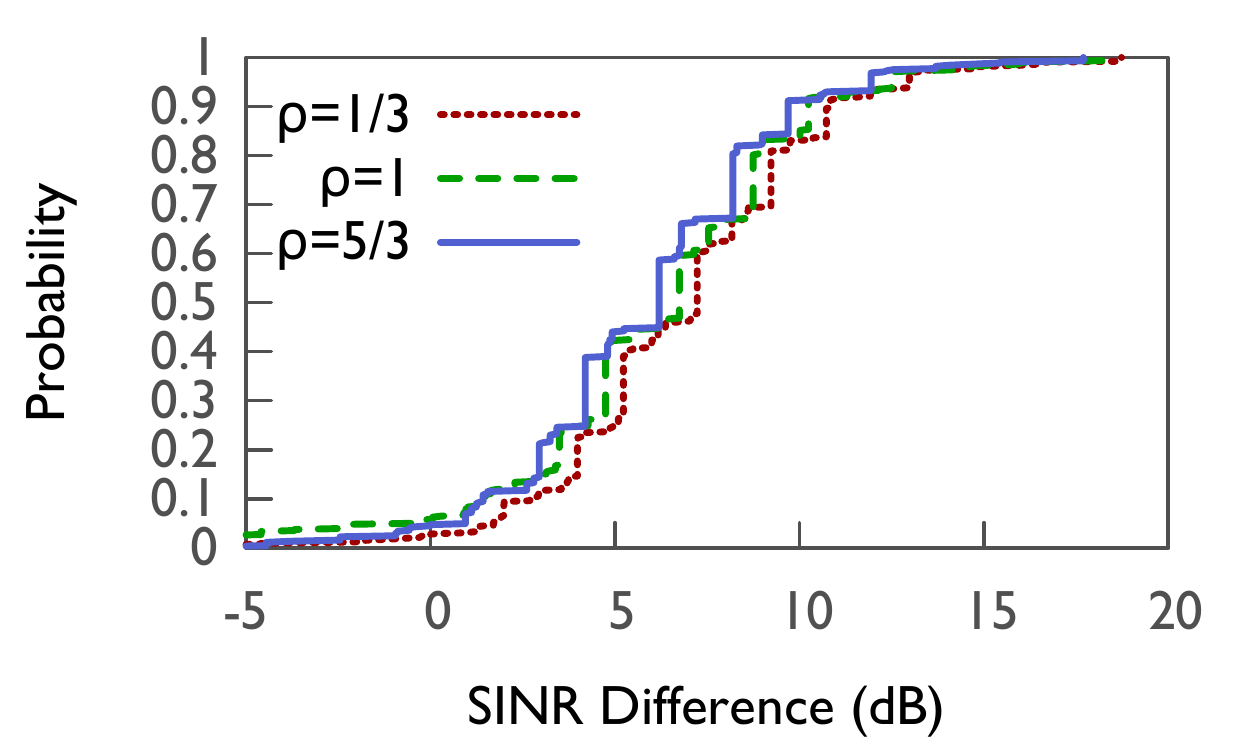}
  \captionof{figure}{SINR gap}
  \label{fig:cqiRsrq}
  \end{subfigure}
  \caption{Factors affecting call drops. Although RSRQ and CQI measures 
      downlink interference, they do not match showing an anomaly.} 
  \end{figure*}

Operators are constantly striving to reduce the amount of drops in the network.
With the move towards carrying voice data also over the data network (VoLTE),
this metric has gained even more importance. This section describes how we used
\NAME{} to analyze drop performance.

\subsubsection{Feature Engineering}
Call drops are normally due to one of three metrics:

\noindent\textbf{Coverage}
It is intuitive to see that poor coverage leads to dropped calls. As seen from
\cref{fig:coverage_cdf}, areas with $RSRP < -130$ dBm have very high connection
drop probability. 

\noindent\textbf{Downlink interference}
For downlink interference, we consider two metrics: RSRQ and downlink CQI. RSRQ
is only reported when the UE might need to handoff. CQI is available independent
of handoffs. From \cref{fig:interference_cdf} and \cref{fig:cqi}, we see that
the distributions do not match. To reveal the difference of these two
distribution, we converted them to the common SINR. To convert CQI, we use the
CQI to SINR table. For RSRQ, we use the formula derived in~\cite{RSRQsinr},
$SINR= \frac{1}{\frac{1}{12RSRQ}-\rho}$. $\rho$ depends on subcarrier
utilization. For two antennas, it is between 1/3 and 5/3. For connection failure
cases, we show the emperical distribution of their SINR differences with 0\%,
50\% and 100\% subcarrier utilization in \cref{fig:cqiRsrq}. We see that 10\%
has a SINR difference of 10 dB. After revealing our finding to experts, it was
discovered that P-CQI feedbacks through physical uplink control channel are not
CRC protected. When spurious P-CQIs are received, the physical link adaptation
algorithm might choose an incorrect rate resulting in drops.

\noindent\textbf{Uplink interference}
As shown in \cref{fig:ul_interference_cdf}, the drop probability for uplink SINR
has a rather steep slope around and peaks at -17dB. The reason is that the
scheduler stops allocating grants at this threshold. 

\subsubsection{Decision Tree Model for Connection Drops} Based on feature
engineering, we picked features that accurately depict call drops. We then used
\NAME{} to train a decision tree that explains the root causes for connection
drops. One of the learned trees is shown in \cref{fig:decision_tree}. As we see,
the tree first classifies drops based on uplink SINR, and then makes use of RSRQ
if available. We confirmed with experts that the model agrees with their
experience. Uplink SINR is more unpredictable because the interference comes
from subscribers associated with neighboring base stations. In contrast,
downlink interference is from neighboring base stations. \NAME{}'s models
achieved an overall accuracy of $92.1\%$ here, while neither a per base station
model nor a global model was able to accurately identify Uplink SINR as the
cause and attained less than $80\%$ accuracy.

\subsubsection{Detecting Cell KPI Change False Positives Using Concept Drift and
    Incremental Learning} An interesting application of \NAME{}'s hybrid model
is in detecting false positives of KPI changes. As explained earlier,
state-of-the-art performance problem detection systems monitor KPIs, and raise
alarms when thresholds are crossed. A major issue with these systems is that the
alarms get raised even for known root causes. However, operators cannot confirm
this without manual investigation resulting in wasted time and effort. This
problem can be solved if known root causes can be filtered out before raising
the alarm. 

\begin{figure*}[!htb]
\centering
\begin{subfigure}{.24\textwidth}
  \centering
  \includegraphics[width=\linewidth]{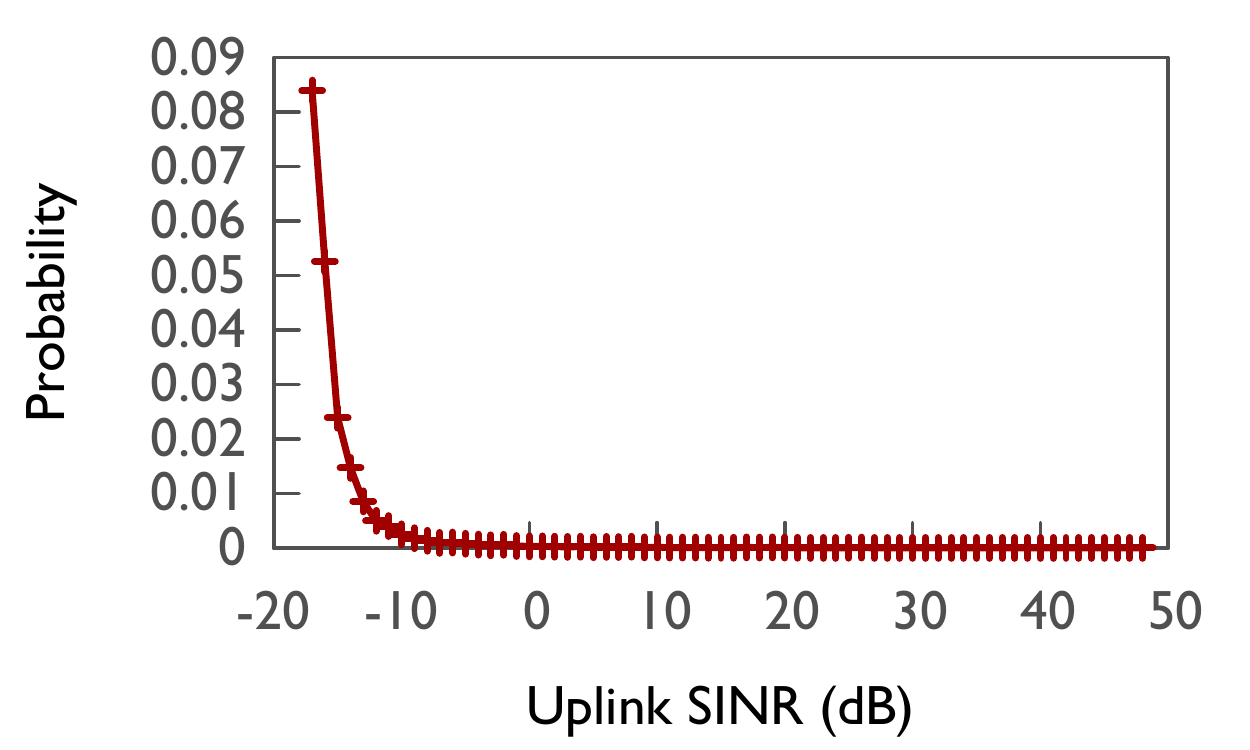}
  \captionof{figure}{Uplink Interference}
  \label{fig:ul_interference_cdf}
  \end{subfigure}%
  \begin{subfigure}{.24\textwidth}
  \centering
  \includegraphics[width=\linewidth]{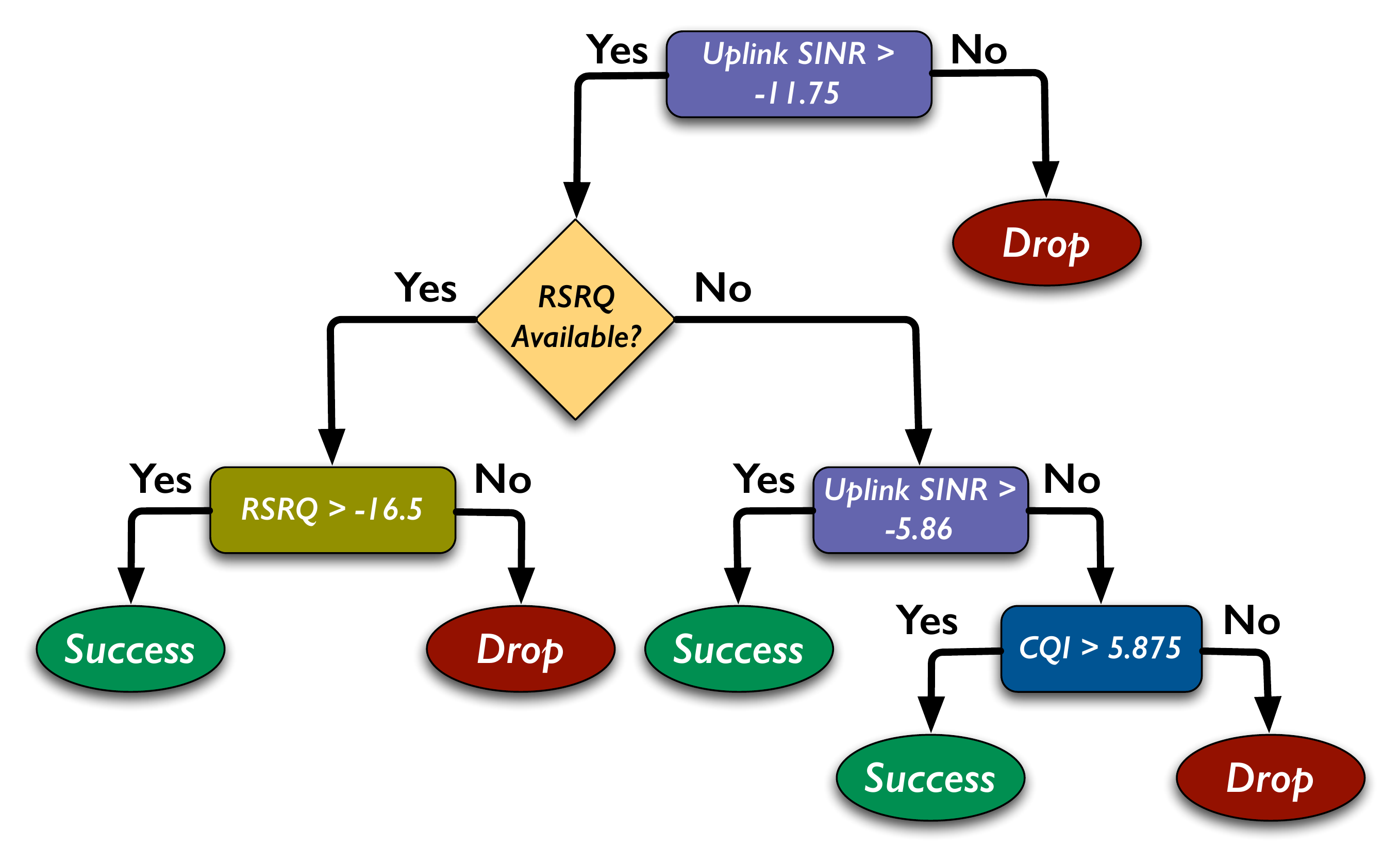}
  \captionof{figure}{Sample decision tree}
  \label{fig:decision_tree}
  \end{subfigure}
  \begin{subfigure}{.24\textwidth}
  \centering
  \includegraphics[width=\linewidth]{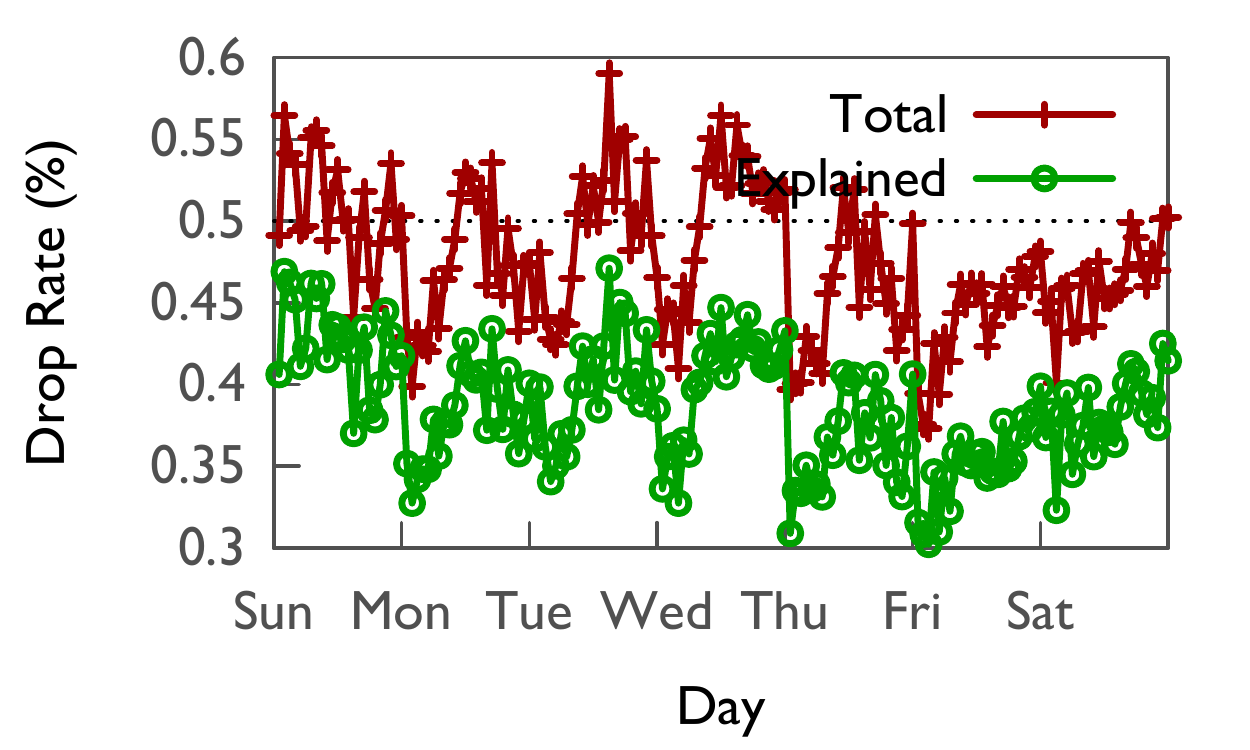}
  \captionof{figure}{Detecting false Alarms}
  \label{fig:falarm}
 \end{subfigure}
 \begin{subfigure}{.24\textwidth}
  \centering
  \includegraphics[width=\linewidth]{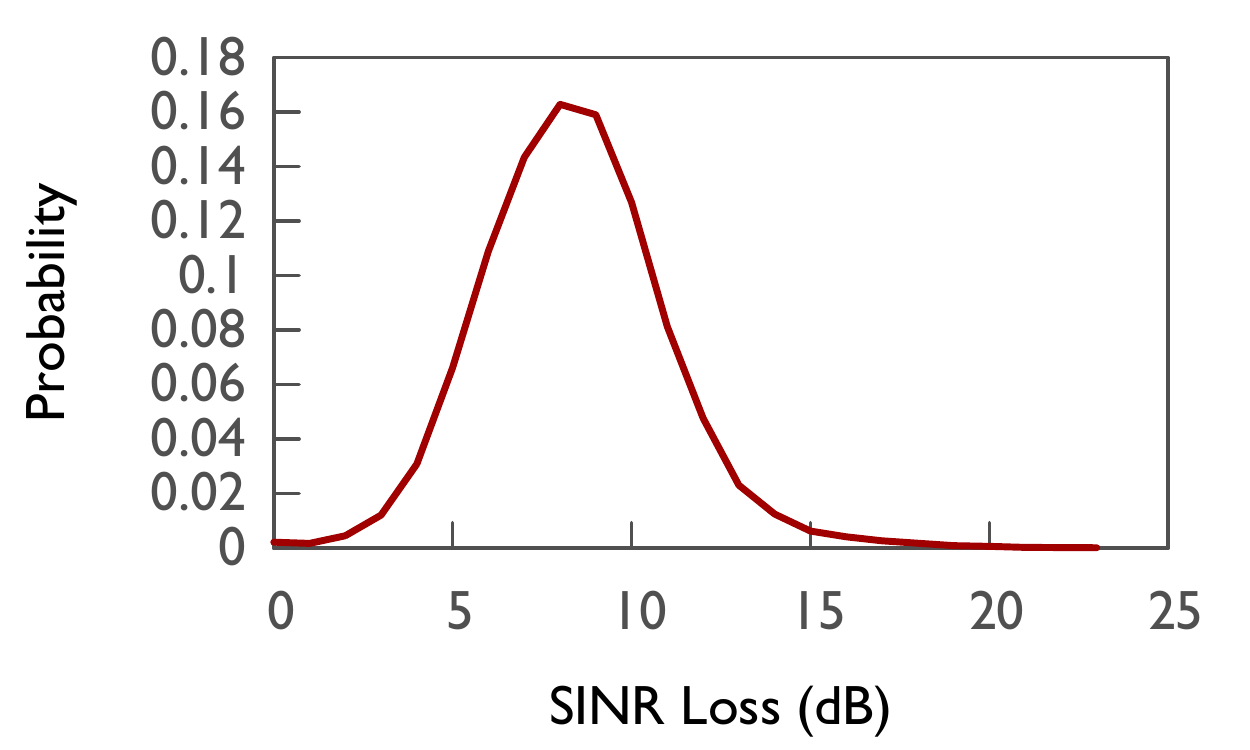}
  \captionof{figure}{Loss of efficiency}
  \label{fig:tput-dbloss}
 \end{subfigure}

 \caption{Uplink interference also affects drops. \NAME{}
     is able to create local models that accurately detects false alarms.}
\end{figure*}

We illustrate this using drop rate. To do so, we use \NAME{} to apply the
decision tree in an incremental fashion on a week worth of data divided into 10
minute interval windows. We used this window length since it matches closely
with an interval that is usually used by the operators for monitoring drop
rates. In every window, we predict the number of drops using our technique. The
predicted drops are explainable, because we know precisely why those drops
happened. We use a threshold of $0.5\%$ for the drop rate, hence anything above
this threshold is marked as an anomaly. The results from this experiment is
depicted in \cref{fig:falarm}. The threshold is exceeded at numerous places.
Normally, these would have to be investigated by the expert. However, \NAME{}
explained them to relieve the burden off the operator.

To estimate the confidence in our prediction, we analyzed our results during the
occurrence of these anomalies. We consider each connection drop or complete
event as a Bernoulli random variable with probability $p$ (from decision tree).
A sequence of $n$ connection events follow a binomial distribution. The 95\%
confidence interval is approximated by $np\pm 2\sqrt{np(1-p)}$. We determine
that the alarm is false if $X$ is within the confidence interval. For this
particular experiment, the bound was found to be (0.7958665,
0.8610155), thus we conclude that \NAME{} was successful.

\subsection{Throughput Performance Analysis}
Our traces report information that lets us compute RLC throughput as ground
truth. We would like to model how far the actual RLC throughput is from the
predicted throughput using physical layer and MAC sub-layer information. This
helps us understand the contributing factors of throughput.  

\subsubsection{Feature Engineering} \textbf{SINR Estimation} The base stations
have two antennas and are capable of MIMO spatial multiplexing (two streams) or
transmit diversity. For both transmissions, each UE reports its two wideband
CQIs. We use the CQI to SINR mapping table used at the base station scheduler to
convert CQI to SINR. For transmission diversity, we convert the two CQIs to a
single SINR as follows. First convert both CQIs to SINR, then compute the two
spectrum efficiencies (bits/sec/Hz) using Shannon capacity. We average the two
spectrum efficiencies and convert it back to SINR. We then add a 3dB
transmission diversity gain to achieve the final SINR. For spatial multiplexing,
we convert the two CQIs to two SINRs.  

\noindent\textbf{Account for PRB control overhead and BLER target} Each PRB is
180 KHz. But not all of it is used for data transmission. For transmit
diversity, a 29\% overhead is incurred per PRB on average because of resources
allocated to physical downlink control channel, broadcast channel and reference
signals. The BLER target is 10\%.

\noindent\textbf{Account for MAC sub-layer retransmissions} The MAC sub-layer
performs retransmissions. We denote the MAC efficiency as $\beta_{MAC}$. It is
computed as the ratio of total first transmissions over total transmissions. Our
traces provide information to compute $\beta_{MAC}$.

\subsubsection{Regression Model: Bearer-Level Throughput} 
The predicted throughput due to transmit diversity is calculated as follows. 
\small
\begin{eqnarray*}
\label{eqn:tput}
tput_{RLCdiv} & = & (1.0-\beta_{MAC}) \times 0.9 \times (1-0.29) \times 180
\times \nonumber \\
& & PRB_{div}  \times log2(1+SINR_{div})/TxTime_{div}
\end{eqnarray*}
\normalsize
$PRB_{div}$ denotes the total PRBs allocated for transmit diversity.
$TxTime_{div}$ is the total transmission time for transmit diversity. Similarly
we can calculate the predicted throughput due to spatial multiplexing. We then
properly weight the two throughput by their respective fraction of transmission
time to derive the final RLC throughput. 

\noindent\textbf{Account for link adaptation in regression model} The base
station does not use the SINR corresponding to the UE reported CQI directly. It
performs link adaptation to achieve the desired BLER of 10\%.  This adaptation
is necessary since the propagation channel is subject to several conditions,
which generally vary in space and time, e.g. path loss, fast fading, UE speed,
location (outdoor, indoor, in-car) etc. We add a variable to the throughput
equation to account for link adaptation, and use \NAME{} to learn it along with
the other unknowns.  Intuitively, this variable indicates how effective the link
adaptation algorithm is.  Since the base station adjusts a maximum of 6dB, we
adjust the SINR used in our prediction by -6dB to compute the lower bound and
+6dB to compute the upper bound. We compute the prediction error as follows. If
the actual throughput is within the two bounds, the error is zero. If the
throughput is outside the two bounds, the error is the distance to the closest
bounds. We characterize the difference between the predicted throughput and
actual throughput in terms of loss in dB. To compute this, we first convert the
actual throughput into SINR. We then subtract the SINR from the one used for
throughput prediction. \Cref{fig:tput-dbloss} shows that the distribution has a
peak around 8dB. As we can see, around 20\% of the bearers have a loss of
efficiency of more than 10 dB. Due to the high fraction of bearers (20\%) with
high dB loss (more than 10 dB), we suspect that the link adaptation algorithm is
slow to adapt to changing conditions. We validate this finding with field
experts. Since this was an unknown insight, they were able to confirm this
observation in their lab tests. The reason for this behavior is because the link
adaptation algorithm uses the moving average SINR, which is a slow mechanism to
adapt.

\section{Discussion} 
\label{sec:discussion}
We have presented a system that resolves the fundamental trade-off between
latency and accuracy in the context of cellular radio network analysis. Now we
discuss the deployability and generality of our solution.

\paragraph{Deployability and Impact:} In several domains, it is common to deploy
research prototypes on a live system to close the loop. Unfortunately, cellular
networks are extremely high business impact systems and hence it is difficult to
deploy our system. Nevertheless, our findings were useful for the operator in
fixing several issues in their network. Moreover, during this experience, we
were able to identify many issues with the data (missing fields, corruptions,
incorrect values, etc) and suggest new fields to be added for better
diagnostics.  \paragraph{Generality:} Our solutions can be classified into two
broad techniques that are applicable to many other domains: partitioning the
data by the underlying cause on which the analysis is to be done; and applying
domain specific formulations to the analysis approach. It is important to note
that MTL is a generic approach. The primary goal of this paper and \NAME{} is
not to apply specific ML algorithms for problem diagnosis, but to propose
approaches that enable accurate applications of any ML technique by resolving
the latency accuracy trade-off. Thus, we believe that our techniques are general
and can be applied to several domains. Some examples of such domains include
spatio-temporal analytics and the Internet-of-Things.

\section{Related Work} 
\label{sec:related} 
\NAME{} is related to cellular network monitoring and
troubleshooting, 
self-organizing networks (SON), and network diagnosis techniques and MTL. 

\textbf{Cellular network monitoring and troubleshooting}
A number of existing cellular network monitoring and diagnosis systems
exist~\cite{GigaScopeSigmod03, WNG, xNetHealing2, NPO}. AT\&T
GigaScope~\cite{GigaScopeSigmod03} and Alcatel-Lucent Wireless Network Guardian
(WNG)~\cite{WNG} generates per IP flow records and monitors many performance
metrics such as aggregate per-cell TCP throughput, delay and loss. Because these
tools tap interfaces in the core networks, they lack information at the radio
access networks. Systems targeting RAN~\cite{xNetHealing2, NPO} typically
monitor aggregate KPIs and per-bearer records separately. Their root cause
analysis of KPI problems correlates with aggregation air interface metrics such
as SINR histograms and configuration data. Because these systems rely on
traditional database technologies, it is hard for them to provide fine-grained
prediction based on per-bearer model. In contrast, \NAME{} is built on top of
efficient big data system, Apache Spark. One recent commercial cellular network
analytics system~\cite{Motive} adopted the Hadoop big data processing framework.
Since it is built on top of WNG~\cite{WNG}, it does not have visibility into
RANs. 
\\
\noindent\textbf{Self-Organizing Networks (SON):}
The goal of SON~\cite{SONLTESpec} is to make the network capable of
self-configuration (e.g. automatic neighbor list configuration) and
self-optimization. \NAME{} focuses on understanding RAN performance and assists
troubleshooting, thus can be used to assist SON. 
\\
\noindent\textbf{Modelling and diagnosis techniques:}
Diagnosing problems in cellular networks has been explored in the literature in
various forms~\cite{LearningParameters, TiresiasICDCS2012, ModelNOMS2008,
eNodeBFaultDetection2006, CellularDependabilityHotdep2013}, where the focus of
the work has either been detecting faults or finding the root cause of failures.
A probabilistic system for auto-diagnosing faults in RAN is presented in
~\cite{LearningParameters}. It relies on KPIs. However, KPIs are not capable of
providing diagnosis at high granularity. Moreover, it is unclear how their
proposals capture complex dependencies between different components in RAN. An
automated approach to locating anomalous events on hierarchical operational
networks was proposed in~\cite{TiresiasICDCS2012} based on hierarchical heavy
hitter based anomaly detection. It is unclear how their proposals carry over to
RAN. Adding autonomous capabilities to alarm based fault detection is discussed
in~\cite{ModelNOMS2008}. While their techniques can help systems auto-heal
faults, correlation based fault detection is insufficient for fine granularity
detection and diagnosis of faults. \cite{eNodeBFaultDetection2006} looks at
detecting call connection faults due to load imbalances.
In~\cite{ArgusInfocom2012}, a technique to detect and localize anomalies from an
ISP point of view is proposed.  Finally,~\cite{CellularDependabilityHotdep2013}
discusses the use of ML tools in predicting call drops and its duration. 
\\
\noindent\textbf{Multi-Task Learning:}
MTL builds on the idea that related tasks can learn from each other to achieve
better statistical
efficiency~\cite{MTLICML1993,MTLNIPS1996,RMLKDD2004,InductiveBiasLearning2000}.
Since the assumption of task relatedness do not hold in many scenarios,
techniques to automatically cluster tasks have been explored in the
past~\cite{TreeMTLICML2010, RobustMTLKDD2012}. However, these techniques
consider tasks as black boxes and hence cannot leverage domain specific
structure. In contrast, \NAME{} proposes a hybrid offline-online MTL formulation
on a domain-specific grouping of tasks based on underlying performance
characteristics.

\section{Conclusion and Future Work}
\label{sec:conclusion}
While several domains can benefit from analyzing data collected in a real-time
fashion, the practicality of these analyses are impeded by a fundamental
trade-off between data collection latency and analysis accuracy. In this paper,
we explored this trade-off in the context of a specific domain use-case,
performance analysis in cellular RANs. We presented \NAME{} to resolve this
trade-off by applying a domain specific formulation of MTL. \NAME{} first
transforms raw data into insightful features. To apply MTL effectively, \NAME{}
proposed a novel PCA inspired similarity metric that groups data from
geographically nearby base stations sharing performance commonalities. Finally,
it also incorporates a hybrid online-offline model for efficient model updates.
We have built \NAME{} on Apache Spark and evaluated it on real data that shows
accuracy improvements ranging from 2.5$\times$ to 4.4$\times$ over direct
applications of ML. We have also used \NAME{} to analyze a live LTE consisting
of over 2 million subscribers for a period of over 10 months, where it uncovered
several problems and insights.

For future work, we wish to explore the applicability of our techniques for
resolving the trade-off in other domains where similarity based grouping is
possible.  Further, since our design is amenable to geo-distributed learning, we
wish to investigate the trade-offs in such settings.

\section*{Acknowledgments}
We thank all AMPLab members who provided feedback on earlier drafts of this paper.
This research is supported in part by NSF CISE Expeditions Award CCF-1139158,
DOE Award SN10040 DE-SC0012463, and DARPA XData Award FA8750-12-2-0331, and
gifts from Amazon Web Services, Google, IBM, SAP, The Thomas and Stacey Siebel
Foundation, Apple Inc., Arimo, Blue Goji, Bosch, Cisco, Cray, Cloudera,
Ericsson, Facebook, Fujitsu, HP, Huawei, Intel, Microsoft, Pivotal, Samsung,
Schlumberger, Splunk, State Farm and VMware.

{\footnotesize 
    \bibliographystyle{acm}
    \bibliography{cellscope}

\begin{thebibliography}{10}

\bibitem{SONLTESpec}
{\sc 3gpp}.
\newblock Self-organizing networks son policy network resource model (nrm)
  integration reference point (irp).
\newblock \url{http://www.3gpp.org/ftp/Specs/archive/32_series/32.521/}.

\bibitem{NetPrintsNSDI2009}
{\sc Aggarwal, B., Bhagwan, R., Das, T., Eswaran, S., Padmanabhan, V.~N., and
  Voelker, G.~M.}
\newblock Netprints: diagnosing home network misconfigurations using shared
  knowledge.
\newblock In {\em Proceedings of the 6th USENIX symposium on Networked systems
  design and implementation\/} (Berkeley, CA, USA, 2009), NSDI'09, USENIX
  Association, pp.~349--364.

\bibitem{WNG}
{\sc {Alcatel Lucent}}.
\newblock 9900 wireless network guardian.
\newblock
  \url{http://www.alcatel-lucent.com/products/9900-wireless-network-guardian},
  2013.

\bibitem{NPO}
{\sc {Alcatel Lucent}}.
\newblock 9959 network performance optimizer.
\newblock
  \url{http://www.alcatel-lucent.com/products/9959-network-performance-optimizer},
  2014.

\bibitem{MotiveSC}
{\sc {Alcatel Lucent}}.
\newblock {Alcatel-Lucent} motive big network analytics for service creation.
\newblock \url{http://resources.alcatel-lucent.com/?cid=170795}, 2014.

\bibitem{Motive}
{\sc {Alcatel Lucent}}.
\newblock Motive big network analytics.
\newblock
  \url{http://www.alcatel-lucent.com/solutions/motive-big-network-analytics},
  2014.

\bibitem{TroubleshootingSigcomm2007}
{\sc Bahl, P., Chandra, R., Greenberg, A., Kandula, S., Maltz, D.~A., and
  Zhang, M.}
\newblock Towards highly reliable enterprise network services via inference of
  multi-level dependencies.
\newblock In {\em Proceedings of the 2007 Conference on Applications,
  Technologies, Architectures, and Protocols for Computer Communications\/}
  (New York, NY, USA, 2007), SIGCOMM '07, ACM, pp.~13--24.

\bibitem{LearningParameters}
{\sc Barco, R., Wille, V., D\'{\i}ez, L., and Toril, M.}
\newblock Learning of model parameters for fault diagnosis in wireless
  networks.
\newblock {\em Wirel. Netw. 16}, 1 (Jan. 2010), 255--271.

\bibitem{InductiveBiasLearning2000}
{\sc Baxter, J.}
\newblock A model of inductive bias learning.
\newblock {\em J. Artif. Int. Res. 12}, 1 (Mar. 2000), 149--198.

\bibitem{MTLICML1993}
{\sc Caruana, R.}
\newblock Multitask learning: A knowledge-based source of inductive bias.
\newblock In {\em Proceedings of the Tenth International Conference on Machine
  Learning\/} (1993), Morgan Kaufmann, pp.~41--48.

\bibitem{CorrelationOSDI2004}
{\sc Cohen, I., Goldszmidt, M., Kelly, T., Symons, J., and Chase, J.~S.}
\newblock Correlating instrumentation data to system states: A building block
  for automated diagnosis and control.
\newblock In {\em Proceedings of the 6th Conference on Symposium on Opearting
  Systems Design \& Implementation - Volume 6\/} (Berkeley, CA, USA, 2004),
  OSDI'04, USENIX Association, pp.~16--16.

\bibitem{GigaScopeSigmod03}
{\sc Cranor, C., Johnson, T., Spataschek, O., and Shkapenyuk, V.}
\newblock Gigascope: a stream database for network applications.
\newblock In {\em Proceedings of the 2003 ACM SIGMOD international conference
  on Management of data\/} (New York, NY, USA, 2003), SIGMOD '03, ACM,
  pp.~647--651.

\bibitem{ensemble}
{\sc Dietterich, T.~G.}
\newblock Ensemble methods in machine learning.
\newblock In {\em Multiple classifier systems}. Springer, 2000, pp.~1--15.

\bibitem{RANvsCore}
{\sc Eljaam, B.}
\newblock Customer satisfaction with cellular network performance: Issues and
  analysis.

\bibitem{xNetHealing1}
{\sc {Ericsson}}.
\newblock {Ericsson} {RAN} analyzer overview.
\newblock \url{http://www.optxview.com/Optimi_Ericsson/RANAnalyser.pdf}, 2012.

\bibitem{xNetHealing2}
{\sc {Ericsson}}.
\newblock {Ericsson} {RAN} analyzer.
\newblock \url{http://www.ericsson.com/ourportfolio/products/ran-analyzer},
  2014.

\bibitem{RMLKDD2004}
{\sc Evgeniou, T., and Pontil, M.}
\newblock Regularized multi--task learning.
\newblock In {\em Proceedings of the Tenth ACM SIGKDD International Conference
  on Knowledge Discovery and Data Mining\/} (New York, NY, USA, 2004), KDD '04,
  ACM, pp.~109--117.

\bibitem{GBT}
{\sc Friedman, J.~H.}
\newblock Greedy function approximation: a gradient boosting machine.
\newblock {\em Annals of statistics\/} (2001), 1189--1232.

\bibitem{ConceptDriftSurvey}
{\sc Gama, J.~a., \v{Z}liobait\.{e}, I., Bifet, A., Pechenizkiy, M., and
  Bouchachia, A.}
\newblock A survey on concept drift adaptation.
\newblock {\em ACM Comput. Surv. 46}, 4 (Mar. 2014), 44:1--44:37.

\bibitem{RobustMTLKDD2012}
{\sc Gong, P., Ye, J., and Zhang, C.}
\newblock Robust multi-task feature learning.
\newblock In {\em Proceedings of the 18th ACM SIGKDD International Conference
  on Knowledge Discovery and Data Mining\/} (New York, NY, USA, 2012), KDD '12,
  ACM, pp.~895--903.

\bibitem{GWPCA}
{\sc Harris, P., Brunsdon, C., and Charlton, M.}
\newblock Geographically weighted principal components analysis.
\newblock {\em International Journal of Geographical Information Science 25},
  10 (2011), 1717--1736.

\bibitem{TiresiasICDCS2012}
{\sc Hong, C.-Y., Caesar, M., Duffield, N., and Wang, J.}
\newblock Tiresias: Online anomaly detection for hierarchical operational
  network data.
\newblock In {\em Proceedings of the 2012 IEEE 32Nd International Conference on
  Distributed Computing Systems\/} (Washington, DC, USA, 2012), ICDCS '12, IEEE
  Computer Society, pp.~173--182.

\bibitem{CellIQNSDI15}
{\sc Iyer, A., Li, L.~E., and Stoica, I.}
\newblock Celliq : Real-time cellular network analytics at scale.
\newblock In {\em 12th USENIX Symposium on Networked Systems Design and
  Implementation (NSDI 15)\/} (Oakland, CA, May 2015), USENIX Association,
  pp.~309--322.

\bibitem{RulebasedDSC2007}
{\sc Khanna, G., Yu~Cheng, M., Varadharajan, P., Bagchi, S., Correia, M.~P.,
  and Ver\'{\i}ssimo, P.~J.}
\newblock Automated rule-based diagnosis through a distributed monitor system.
\newblock {\em IEEE Trans. Dependable Secur. Comput. 4}, 4 (Oct. 2007),
  266--279.

\bibitem{TreeMTLICML2010}
{\sc Kim, S., and Xing, E.~P.}
\newblock Tree-guided group lasso for multi-task regression with structured
  sparsity.
\newblock {\em Intenational Conference on Machine Learning (ICML)\/} (2010).

\bibitem{MLbase13}
{\sc Kraska, T., Talwalkar, A., Duchi, J.~C., Griffith, R., Franklin, M.~J.,
  and Jordan, M.~I.}
\newblock Mlbase: A distributed machine-learning system.
\newblock In {\em CIDR\/} (2013).

\bibitem{PCASimilarity}
{\sc Krzanowski, W.}
\newblock Between-groups comparison of principal components.
\newblock {\em Journal of the American Statistical Association 74}, 367 (1979),
  703--707.

\bibitem{AnomaliesSigcomm2004}
{\sc Lakhina, A., Crovella, M., and Diot, C.}
\newblock Diagnosing network-wide traffic anomalies.
\newblock In {\em Proceedings of the 2004 Conference on Applications,
  Technologies, Architectures, and Protocols for Computer Communications\/}
  (New York, NY, USA, 2004), SIGCOMM '04, ACM, pp.~219--230.

\bibitem{ModelNOMS2008}
{\sc Liu, Y., Zhang, J., Jiang, M., Raymer, D., and Strassner, J.}
\newblock A model-based approach to adding autonomic capabilities to network
  fault management system.
\newblock In {\em Network Operations and Management Symposium, 2008. NOMS 2008.
  IEEE\/} (April 2008), pp.~859--862.

\bibitem{ML}
{\sc Murphy, K.~P.}
\newblock {\em Machine learning: a probabilistic perspective}.
\newblock MIT press, 2012.

\bibitem{ScalableMTL}
{\sc Pan, Y., Xia, R., Yin, J., and Liu, N.}
\newblock A divide-and-conquer method for scalable robust multitask learning.
\newblock {\em Neural Networks and Learning Systems, IEEE Transactions on 26},
  12 (Dec 2015), 3163--3175.

\bibitem{PCA}
{\sc Pearson, K.}
\newblock {On lines and planes of closest fit to systems of points in space}.
\newblock {\em Philosophical Magazine 2}, 6 (1901), 559--572.

\bibitem{eNodeBFaultDetection2006}
{\sc Rao, S.}
\newblock Operational fault detection in cellular wireless base-stations.
\newblock {\em IEEE Trans. on Netw. and Serv. Manag. 3}, 2 (Apr. 2006), 1--11.

\bibitem{4G}
{\sc Safwat, A.~M., and Mouftah, H.}
\newblock 4g network technologies for mobile telecommunications.
\newblock {\em Network, IEEE 19}, 5 (2005), 3--4.

\bibitem{RSRQsinr}
{\sc Salo, J.}
\newblock Mobility parameter planning for {3GPP LTE}: Basic concepts and
  intra-layer mobility.
\newblock \url{www.lteexpert.com/lte_mobility_wp1_10June2013.pdf}.

\bibitem{LTE}
{\sc Sesia, S., Toufik, I., and Baker, M.}
\newblock {\em LTE: the UMTS long term evolution}.
\newblock Wiley Online Library, 2009.

\bibitem{NetworkDynamicsSigmetrics2014}
{\sc Shafiq, M.~Z., Erman, J., Ji, L., Liu, A.~X., Pang, J., and Wang, J.}
\newblock Understanding the impact of network dynamics on mobile video user
  engagement.
\newblock In {\em The 2014 ACM International Conference on Measurement and
  Modeling of Computer Systems\/} (New York, NY, USA, 2014), SIGMETRICS '14,
  ACM, pp.~367--379.

\bibitem{PerformanceSigmetrics2013}
{\sc Shafiq, M.~Z., Ji, L., Liu, A.~X., Pang, J., Venkataraman, S., and Wang,
  J.}
\newblock A first look at cellular network performance during crowded events.
\newblock In {\em Proceedings of the ACM SIGMETRICS/International Conference on
  Measurement and Modeling of Computer Systems\/} (New York, NY, USA, 2013),
  SIGMETRICS '13, ACM, pp.~17--28.

\bibitem{L1SML2011}
{\sc Shalev-Shwartz, S., and Tewari, A.}
\newblock Stochastic methods for l1-regularized loss minimization.
\newblock {\em J. Mach. Learn. Res. 12\/} (July 2011), 1865--1892.

\bibitem{3G}
{\sc Smith, C.}
\newblock {\em 3G wireless networks}.
\newblock McGraw-Hill, Inc., 2006.

\bibitem{MLI13}
{\sc Sparks, E.~R., Talwalkar, A., Smith, V., Kottalam, J., Pan, X., Gonzalez,
  J.~E., Franklin, M.~J., Jordan, M.~I., and Kraska, T.}
\newblock {MLI:} an {API} for distributed machine learning.
\newblock In {\em 2013 {IEEE} 13th International Conference on Data Mining,
  Dallas, TX, USA, December 7-10, 2013\/} (2013), H.~Xiong, G.~Karypis, B.~M.
  Thuraisingham, D.~J. Cook, and X.~Wu, Eds., {IEEE} Computer Society,
  pp.~1187--1192.

\bibitem{3gpp-specs}
{\sc {Technical Specification Group}}.
\newblock 3gpp specifications.
\newblock \url{http://www.3gpp.org/specifications}.

\bibitem{CellularDependabilityHotdep2013}
{\sc Theera-Ampornpunt, N., Bagchi, S., Joshi, K.~R., and Panta, R.~K.}
\newblock Using big data for more dependability: A cellular network tale.
\newblock In {\em Proceedings of the 9th Workshop on Hot Topics in Dependable
  Systems\/} (New York, NY, USA, 2013), HotDep '13, ACM, pp.~2:1--2:5.

\bibitem{MTLNIPS1996}
{\sc Thrun, S.}
\newblock Is learning the n-th thing any easier than learning the first?
\newblock In {\em Advances in Neural Information Processing Systems\/} (1996),
  The MIT Press, pp.~640--646.

\bibitem{Lasso1994}
{\sc Tibshirani, R.}
\newblock Regression shrinkage and selection via the lasso.
\newblock {\em Journal of the Royal Statistical Society, Series B 58\/} (1994),
  267--288.

\bibitem{PeerpressureOSDI2004}
{\sc Wang, H.~J., Platt, J.~C., Chen, Y., Zhang, R., and Wang, Y.-M.}
\newblock Automatic misconfiguration troubleshooting with peerpressure.
\newblock In {\em Proceedings of the 6th Conference on Symposium on Opearting
  Systems Design \& Implementation - Volume 6\/} (Berkeley, CA, USA, 2004),
  OSDI'04, USENIX Association, pp.~17--17.

\bibitem{ArgusInfocom2012}
{\sc Yan, H., Flavel, A., Ge, Z., Gerber, A., Massey, D., Papadopoulos, C.,
  Shah, H., and Yates, J.}
\newblock Argus: End-to-end service anomaly detection and localization from an
  isp's point of view.
\newblock In {\em INFOCOM, 2012 Proceedings IEEE\/} (2012), pp.~2756--2760.

\bibitem{pcasimilaritytimevariate}
{\sc Yang, K., and Shahabi, C.}
\newblock A pca-based similarity measure for multivariate time series.
\newblock In {\em Proceedings of the 2nd ACM International Workshop on
  Multimedia Databases\/} (New York, NY, USA, 2004), MMDB '04, ACM, pp.~65--74.

\bibitem{RDD}
{\sc Zaharia, M., Chowdhury, M., Das, T., Dave, A., Ma, J., McCauley, M.,
  Franklin, M.~J., Shenker, S., and Stoica, I.}
\newblock Resilient distributed datasets: a fault-tolerant abstraction for
  in-memory cluster computing.
\newblock In {\em Proceedings of the 9th USENIX conference on Networked Systems
  Design and Implementation\/} (Berkeley, CA, USA, 2012), NSDI'12, USENIX
  Association, pp.~2--2.

\bibitem{SparkStreamingSOSP13}
{\sc Zaharia, M., Das, T., Li, H., Hunter, T., Shenker, S., and Stoica, I.}
\newblock Discretized streams: Fault-tolerant streaming computation at scale.
\newblock In {\em Proceedings of the Twenty-Fourth ACM Symposium on Operating
  Systems Principles\/} (New York, NY, USA, 2013), SOSP '13, ACM, pp.~423--438.

\bibitem{FeatureSelectionSIGMOD2014}
{\sc Zhang, C., Kumar, A., and R{\'e}, C.}
\newblock Materialization optimizations for feature selection workloads.
\newblock In {\em Proceedings of the 2014 ACM SIGMOD International Conference
  on Management of Data\/} (New York, NY, USA, 2014), SIGMOD '14, ACM,
  pp.~265--276.

\bibitem{FailureICAC2004}
{\sc Zheng, A.~X., Lloyd, J., and Brewer, E.}
\newblock Failure diagnosis using decision trees.
\newblock In {\em Proceedings of the First International Conference on
  Autonomic Computing\/} (Washington, DC, USA, 2004), ICAC '04, IEEE Computer
  Society, pp.~36--43.

\end{thebibliography}
}

\end{document}